\documentclass[11pt]{article}
\usepackage{amsmath,amssymb,mathtools}
\usepackage{acl}
\usepackage{colortbl}
\usepackage{bm}
\usepackage{xcolor}
\usepackage{array}
\usepackage{booktabs}
\usepackage{multirow}
\usepackage{adjustbox}
\usepackage{makecell}
\usepackage{fontawesome}     
\usepackage{pifont}
\usepackage{graphicx}
\usepackage{makecell}
\usepackage{array}
\usepackage{float}   
\usepackage{placeins}
\usepackage{enumitem}
\usepackage{tcolorbox}

\newcommand{\yes}{\raisebox{-0.1em}{\includegraphics[height=0.75em]{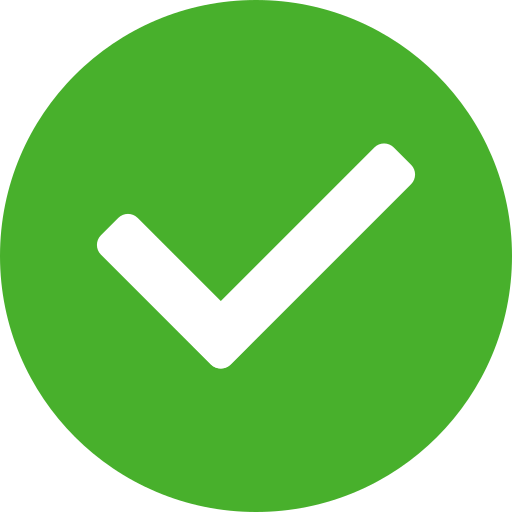}}}
\newcommand{\no}{\raisebox{-0.1em}{\includegraphics[height=0.75em]{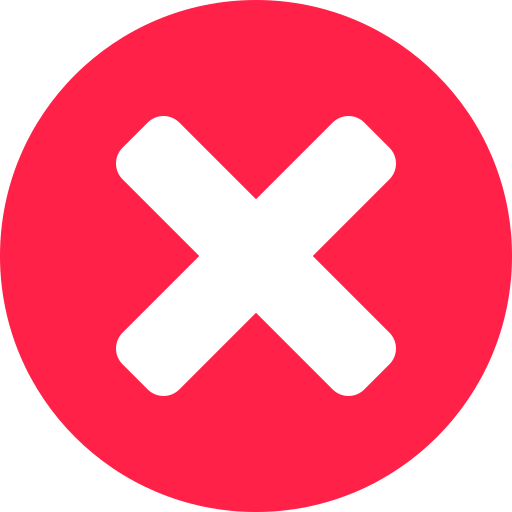}}}
\newcommand{\first}{\raisebox{-0.1em}{\includegraphics[height=1.1em]{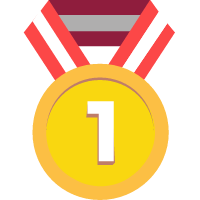}}}
\newcommand{\second}{\raisebox{-0.1em}{\includegraphics[height=1.1em]{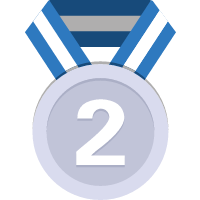}}}
\usepackage{times}
\usepackage{latexsym}

\usepackage[T1]{fontenc}

\usepackage[utf8]{inputenc}

\usepackage{microtype}

\usepackage{inconsolata}

\usepackage{graphicx}

\usepackage{booktabs}       
\usepackage{multirow}       
\usepackage{graphicx}       
\usepackage[table]{xcolor}  
\usepackage{amssymb}       
\usepackage{array}         
\usepackage{adjustbox}
\usepackage{makecell}
\usepackage{enumitem}
\usepackage{threeparttable}
\definecolor{headergray}{RGB}{235, 236, 240}

\usepackage{booktabs} 
\usepackage{array}    
\usepackage{tabularx}  

\title{ChronosAudio: A Comprehensive Long-Audio Benchmark for Evaluating Audio-Large Language Models}

\author{
 \textbf{Kaiwen Luo\textsuperscript{\rm 1,*}},
 \textbf{Liang Lin\textsuperscript{\rm 1,*}},
 \textbf{Yibo Zhang\textsuperscript{\rm 2,*}},
 \textbf{Moayad Aloqaily\textsuperscript{3}},
 \textbf{Jialiang Tao\textsuperscript{4}},
 \textbf{Dexian Wang\textsuperscript{5}},
 \\
 \textbf{Zhenhong Zhou\textsuperscript{1}},
 \textbf{Junwei Zhang\textsuperscript{2}},
 \textbf{Kun Wang \textsuperscript{\rm 1,\dag}},
 \textbf{Li Sun\textsuperscript{\rm 4,\dag}},
 \textbf{Qingsong Wen\textsuperscript{6}}
\\
 \textsuperscript{1}Nanyang Technological University,
 \\
 \textsuperscript{2}Independent Researcher,
 \\
 \textsuperscript{3}United Arab Emirates University,
 \\
 \textsuperscript{4}North China Electric Power University,
 \\
 \textsuperscript{5}Southwest Jiaotong University,
 \\
 \textsuperscript{6}Squirrel AI Learning 
\\
 \small{
   \textbf{Correspondence:} \href{mailto:email@domain}{kaiwenluo74@gmail.com}
 }
}

\begin{document}
\maketitle
\begin{abstract}
Although Audio Large Language Models (ALLMs) have witnessed substantial advancements, their long audio understanding capabilities remain unexplored. Existing benchmarks have been proposed for general audio tasks, they predominantly focus on short clips, leaving without a consensus on evaluating ALLMs over extended durations. This paper proposes \textit{ChronosAudio}, the first multi-task benchmark tailored for long-audio understanding in ALLMs. It encompasses six task categories and comprises 36,000 test instances totaling over 200 hours audio, stratified into short, middle, and long-form categories to comprehensively evaluate length generalization. Extensive experiments on 16 state-of-the-art models using \textit{ChronosAudio} yield three critical findings: \ding{182} \textbf{Precipitous Long-Context Collapse}: ALLMs exhibit a severe inability to sustain performance, with the transition from short to long contexts triggering a staggering performance degradation of over 90\% in specific tasks. \ding{183} \textbf{Structural Attention Dilution}: Performance degradation stems from a fundamental failure in maintaining temporal locality; attention mechanisms suffer from significant diffusion in later sequences. \ding{184} \textbf{Restorative Ceiling of Mitigation}: Current strategies only offer 50\% recovery. These findings reveal significant challenges in long-audio, underscoring the urgent need for approaches to achieve robust, document-level audio reasoning. The codes and dataset are available at \url{https://anonymous.4open.science/r/ChronosAudio-D49A}
\end{abstract}

\section{Introduction}\label{sec:intro}

Recent advancements in Large Language Models (LLMs) have catalyzed the evolution of Multimodal LLMs (MLLMs), significantly enhancing their capabilities in audio understanding \citep{wang2025comprehensive}, instruction following \citep{feng2025follow}, and reasoning \citep{lu2025auditing}. Building upon these foundations, state-of-the-art Audio Large Language Models (ALLMs) \citep{an2024qwen2,fan2025qwen25,wu2025step,Ding2025kimi, luo2026survey} have emerged to handle various audio-related tasks \citep{huang2023speech,sun2025spatial,tian2025ualm,chen2025audio}.
\begin{figure*}[t] 
    \centering
    \includegraphics[width=\linewidth]{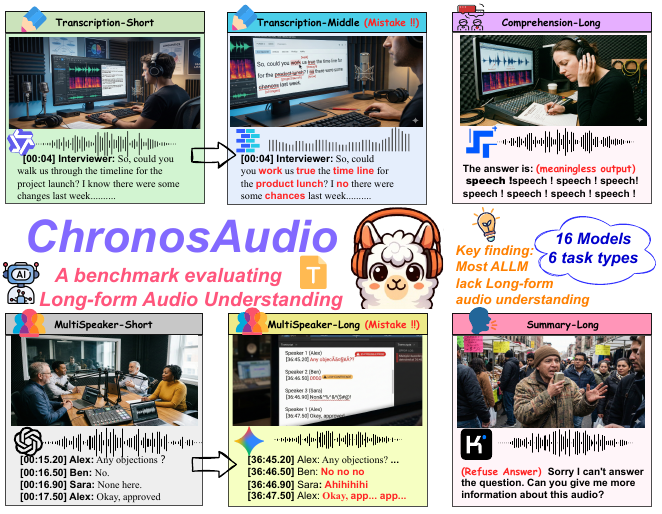}
    \caption{An overview of ChronosAudio provides a multi-dimensional framework for systematically ALLMs.}
    \label{fig:introduction} 
\end{figure*} 

\begin{table}[t]
\vspace{-1.0em}
\centering

\definecolor{colModel}{gray}{0.96}   %
\definecolor{colBlue}{HTML}{F0F6FF}  
\definecolor{colOrng}{HTML}{FFF4E6} 
\definecolor{colTeal}{HTML}{E8FDF5}  
\definecolor{colPurp}{HTML}{F3E5F5}

\renewcommand{\arraystretch}{1.15} 
\setlength{\tabcolsep}{1pt} 
\renewcommand\theadfont{\bfseries} 

\begin{adjustbox}{width=1\linewidth}
\huge 

\begin{tabular}{
    >{\columncolor{colModel}}l | 
    >{\columncolor{colBlue}}c | 
    >{\columncolor{colOrng}}c | 
    >{\columncolor{colTeal}}c | 
    >{\columncolor{colPurp}}c
} 
    \hline

    \textbf{Benchmark} & 
    \thead{Multi\\Dim.} & 
    \thead{Long\\Task} & 
    \thead{Total\\Dur.} & 
    \thead{Avg.\\Dur.} \\
    \hline

    AudioBench \citep{wang2023audiobench} & \yes & \no & 400h & 14s \\
    \hline
     
    AIR-Bench \citep{yang2024airbench} & \yes & \no & 252h & 35s \\
    \hline

    MuChoMusic \citep{weck2024muchomusic} & \yes & \no & 5h & 26s \\
    \hline
        
    MMSU \citep{wang2025mmsu} & \yes & \no & 10h & 7s \\
    \hline
        
    MMAU \citep{sakshi2025mmau} & \yes & \no & 28h & 10s \\
    \hline
     
    MMAR \citep{ma2025mmar} & \yes & \no & 6h & 19s  \\
    \hline
     
    BLAB \citep{ahia2025blab} & \no & \yes & 833h & 51m  \\
    \hline

    AudioMarathon \citep{he2025audiomarathon} & \no & \yes & 392h & 213s  \\
    \hline

    \textbf{ChronosAudio (Ours)} & \yes & \yes & 237h & 322s \\
    \hline
\end{tabular}
\end{adjustbox}
\caption{Comparison between \textit{ChronosAudio} with other audio benchmarks.}
\label{table:bench}
\vspace{-1em}
\end{table}

The burgeoning field of ALLMs has spurred the emergence of numerous datasets \citep{li2025isabench, cheng2025ahabench, lin2025hidden,li2025audiorole} aimed at broadly evaluating their general capabilities. Representative benchmarks include Dynamic-SUPERB \citep{tseng2023dynamic}, Audio-Bench \citep{wang2023audiobench}, and AIR-Bench \citep{yang2024airbench}, alongside those prioritizing safety and trustworthiness like AudioTrust \citep{li2025audiotrust} and Audio Jailbreak \citep{song2025audio}. However, Table \ref{table:bench} shows that their data is confined to short audio \citep{li2025audiorole,zhang2025wildspeech}, typically spanning merely seconds to a few minutes. While the recently proposed BLAB \cite{ahia2025blab} and AudioMarathon \cite{he2025audiomarathon} pioneer raw stream length, they primarily behave as a massive monolithic diagnostic. In contrast, \textit{ChronosAudio} provides a highly multi-dimensional task matrix. Now systematic research into the long-duration audio understanding and analysis capabilities of ALLMs remains largely unexplored. Consequently, this raises a natural question: \textit{How do Audio Large Language Models perform on long audio?}

To explore this question systematically, we introduce \textit{ChronosAudio}, the first multi-task benchmark tailored for long-context evaluating \citep{chen2023longlora, ding2024longrope,li2024longcontext,gao2025nextlong} on ALLMs as shown in Figure \ref{fig:introduction}. \textit{ChronosAudio} is composed of \textbf{six major task categories, covering key long-audio application scenarios including Dictation, Localization, Transcription, Multi-Speaker, Comprehension, and Summary}. In addition, ChronosAudio contains \textbf{3,000} short-form audios (30s$\sim$5min), \textbf{2,000} middle-form audios (5min$\sim$10min) and \textbf{1,000} long-form audios (10min$\sim$20min) to provide a more comprehensive evaluation of the ALLMs' capabilities on long contexts. ChronosAudio includes \textbf{36,000} test instances, over \textbf{200} h audio, and an average length of \textbf{about 6} min.

Building on the \textit{ChronosAudio} benchmark, our extensive experimentation rigorously demonstrates that state-of-the-art ALLMs \textbf{exhibit a critical vulnerability in maintaining high-fidelity understanding when processing extended temporal contexts.} 
For instance, we observe a precipitous performance collapse across evaluated models, where the transition from short to long contexts triggers a staggering degradation of over 90\% in specific tasks. 
Even top-tier proprietary models are not immune to this decline; while closed-source models achieve a Transcription score of 42.66 on short audio, this metric plummets to a mere 3.86 when processing long-form sequences. 
However, a distinct performance chasm is observed between model categories: while closed-source models maintain robust reasoning capabilities with a Summary score of 59.22 on long audio, open-source counterparts suffer a structural collapse to an average score of 14.14 in the same scenarios.

\noindent \textbf{Experimental Takeaways.}
\begin{itemize}[noitemsep, topsep=0pt, parsep=0pt, leftmargin=*]
    \item \textbf{Precipitous Long-Context Collapse.}
    ALLMs exhibit a severe inability to sustain performance over extended durations. Our evaluation reveals that the transition from short to long contexts triggers a staggering performance degradation of over \textbf{90\%} in specific tasks, with the \textit{Lost-in-the-Middle} phenomenon manifesting significantly earlier than in other textual models.

    \item \textbf{Structural Attention Dilution.}
    Performance degradation stems from a fundamental failure in maintaining temporal locality. Visualization of attention weights confirms that while models capture local dependencies in early stages, the attention mechanism suffers from significant diffusion in later sequences, resulting in a loss of high-resolution alignment.

    \item \textbf{Restorative Ceiling of Mitigation.}
    Current long-context strategies offer limited recovery. While mechanisms like Sparse Attention improve stability, they encounter a ceiling, recovering only \textbf{50\%} of the model's short-context proficiency, highlighting simple architectural tweaks are insufficient for long audio reasoning.
\end{itemize}
\section{Preliminary \& Related Work}\label{sec:related}

\subsection{Audio Large Language Models}
\label{sec:related_audio_llm}

Recent ALLMs have demonstrated significant potential in unifying tasks such as automatic speech recognition \citep{min2023exploring, bai2024seed, xie2025thinking, carone2025muse}, speech translation \citep{huang2023speech, liu2025tta}, audio generation \citep{Wang2025SparkTTSAE}, and dialogue \citep{zhang2023speechgpt, gong2023listen}.

ALLMs typically employ a two stage paradigm that aligns continuous audio signals with the discrete textual space of LLMs. In the first stage, the continuous audio waveform $x_a \in \mathbb{R}^{T_a}$ is discretized into a sequence of acoustic tokens to bridge the modality gap. This is generally achieved via a neural audio tokenizer or codec $\phi_a$, which maps the input to a discrete codebook space $\mathbb{Z}^{L_a}$:
\begin{equation}
\mathbf{c}_a = \phi_a(x_a), \quad \phi_a : \mathbb{R}^{T_a} \rightarrow \mathbb{Z}^{L_a},
\label{eq:audio_tokenizer}
\end{equation}
where $L_a$ denotes the length of result tokens. 

To enable multimodal interaction, these acoustic tokens are projected into a shared semantic embedding space alongside textual inputs. Specifically, given a textual prompt $x_t \in \mathbb{Z}^{L_t}$, the model constructs a unified input sequence $\mathbf{z}$ by concatenating the embeddings of both modalities:
\begin{equation}
\mathbf{z} = [\phi_e(\mathbf{c}_a) \mathbin{\Vert} \phi_t(x_t)] \in \mathbb{R}^{(L_a+L_t)\times d},
\label{eq:concat}
\end{equation}
where $\phi_e$ and $\phi_t$ serve as the embedding functions and $\Vert$ denotes the concatenation operation.

The unified embedding sequence is subsequently processed by a Transformer decoder $f_\theta$, which models the probability distribution over a joint vocabulary $\mathcal{V}$ containing textual and acoustic units. $\mathbf{h} = f_\theta(\mathbf{z})$ are projected via a matrix $W \in \mathbb{R}^{|\mathcal{V}| \times d}$ to compute the next-token probabilities:
\begin{equation}
\hat{y} = \operatorname{softmax}(W\mathbf{h}).
\label{eq:softmax}
\end{equation}
Eq \ref{eq:softmax} often referred to as joint autoregressive decoding, allows ALLMs to coherently generate interleaved audio and textual outputs, thereby facilitating complex tasks such as audio captioning \citep{guo2025brace} and dialogue \citep{gosai2025audio}.

\subsection{Long Context Understanding}
\label{sec:lcc_llms}

In the realm of text-modal LLMs, long-context understanding, defined as the capacity to process and reason over extended sequences, has witnessed systematic advancements. Recent work has examined how LLMs process very long contexts\cite{liu2024lost, wang2024adaleval, kuratov2024babilong, wu2024longgenbench, yang2025rethinking}. These studies include stress tests for retrieval in book-length documents, where performance degrades when key information is located in the middle part\cite{bai2024longbench, bai2024longbenchv2, que2024hellobench}. These works have established a solid foundation for evaluating retrieval and reasoning in long-text scenarios.

Despite advancements in NLP, research on long-context understanding in the audio modality remains scarce, primarily because audio is continuous, making reasoning over long sequential structures, significantly more challenging than with discrete text tokens \citep{zhu2025fastlongspeech}. Furthermore, current sota ALLMs lack sufficient benchmarking for these long scenarios, as existing evaluations predominantly focus on short clips \citep{ahia2025blab, ma2025mmar, wang2025mmsu, sakshi2025mmau, li2025isa}. This highlights a critical need for dedicated benchmarks and modeling approaches that target long-context understanding.
\section{ChronosAudio}
\label{sec:methodology}

We introduce \textit{ChronosAudio}, a multi-task benchmark specifically designed to assess long-audio understanding on \textbf{16} sota ALLMs. The benchmark organizes six tasks into three distinct categories: \textit{Temporal Perception}, \textit{Verbatim Sequential Generation}, and \textit{High-level Reasoning}.

\subsection{An Overview of ChronosAudio}
\label{subsec:dataset}
To ensure a comprehensive evaluation of length generalization and address the scarcity of long-context benchmarks, \textit{ChronosAudio} explicitly focuses on time-extended audio scenarios. The benchmark aggregates a diverse collection of audio data, totally amounting to over \textbf{200} hours. 

\textit{ChronosAudio} is constructed as a purely synthetic benchmark SQuAD \citep{yatskar2019qualitative}. The corresponding audio was generated using the F5-TTS engine \citep{chen2025f5}.

Specifically, ChronosAudio comprises \textbf{36,000} test instances derived from \textbf{6,000} distinct audio clips. These instances span six major task categories, ensuring broad coverage of key long-context application scenarios. The audio data is rigorously stratified into three duration categories to probe model capabilities across different context windows. We divide data into three types:
\begin{itemize}[noitemsep, topsep=0pt, parsep=0pt, leftmargin=*]
    \item \textbf{Short-form:} 3,000 audio clips (30s$\sim$5min), serving as a baseline for context handling.
    \item \textbf{Middle-form:} 2,000 audio clips (5min$\sim$10min), bridging the gap between sentence-level and document-level processing.
    \item \textbf{Long-form:} 1,000 audio clips (10min$\sim$20min), specifically designed to challenge the limits of long-context attention and reasoning.
\end{itemize}

\begin{figure}[t!]
    \centering
    \makebox[\linewidth][c]{
        \includegraphics[width=\linewidth]{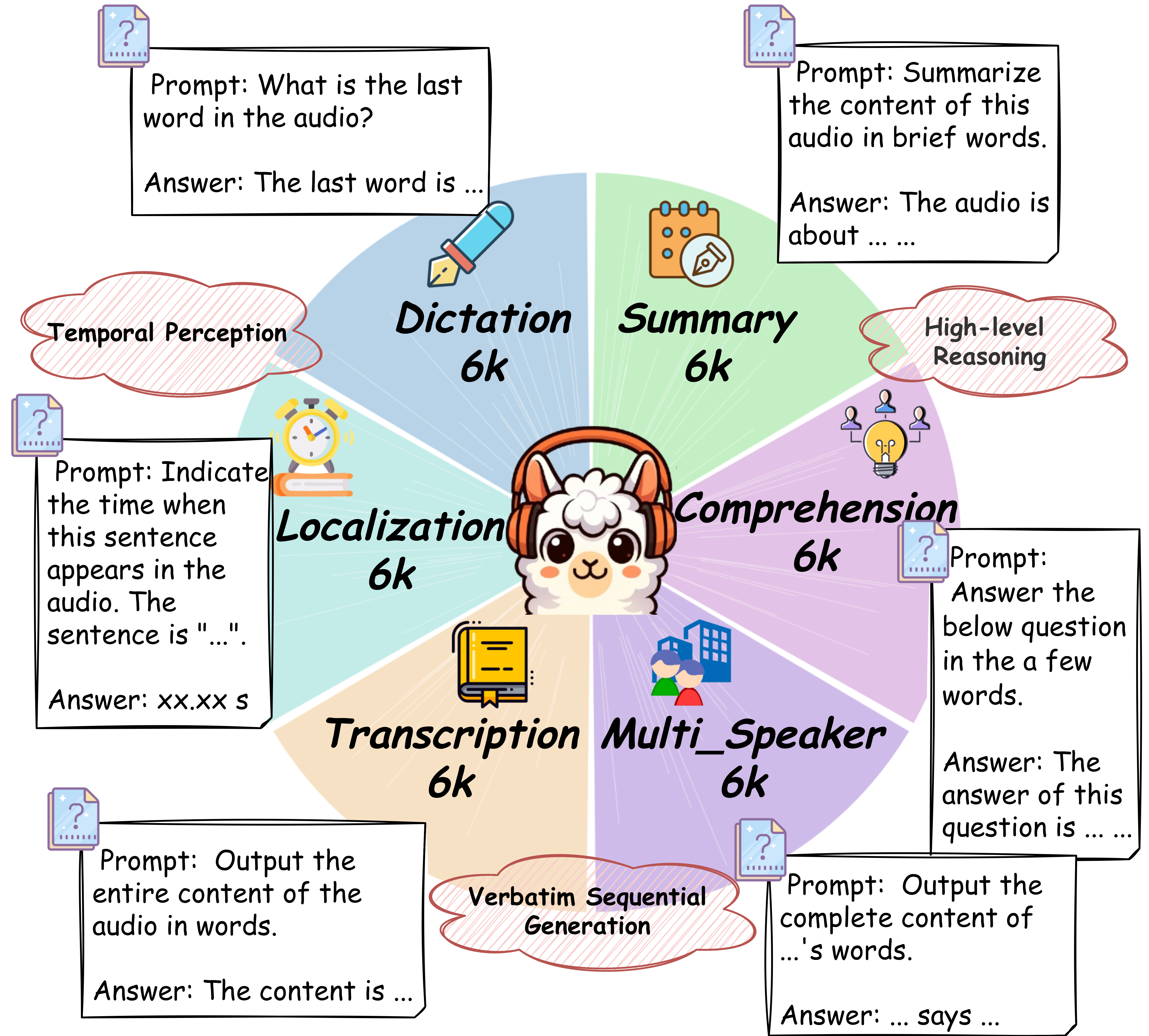}
    }
    \vspace{-1.0em}
    \caption{Overview of the ChronosAudio benchmark data composition. Each task comprises 6,000 test instances, totaling 36,000 samples.}
    \label{fig:task}
\end{figure}

Figure \ref{fig:task} shows that all datasets in ChronosAudio are standardized into a unified format. This standardization is critical to facilitate consistent, automated evaluation across diverse audio-centric capabilities and to enable fair comparisons between different performance of ALLMs.

\subsection{Temporal Perception}
\label{subsec:temporal}
This category focuses on the model's ability to locate specific information in the time domain and sustain attention over long durations.

\noindent \textbf{Dictation.}
This task probes attention persistence. The metric, Dictation Accuracy ($\mathrm{Acc}_{\text{dict}}$), is calculated as the ratio of correctly transcribed units ($\mathrm{N}_{\text{correct}}$) to total reference units ($\mathrm{N}_{\text{total}}$):
\begin{equation}
    \mathrm{DS} = \frac{\mathrm{N}_{\text{correct}}}{\mathrm{N}_{\text{total}}},
\end{equation}
where $\mathrm{N}_{\text{correct}}$ denotes the count of units (words or characters) correctly transcribed by the model, and $\mathrm{N}_{\text{total}}$ represents the total number of units in the reference text.

\noindent \textbf{Localization.}
This task measures temporal grounding. We first compute the absolute temporal error $\Delta \mathrm{t} = |\mathrm{t}_{\text{pred}} - \mathrm{t}_{\text{ref}}|$. With a strict tolerance threshold $\tau = 0.1\text{s}$, the localization score is defined as:
\begin{equation}
    \mathrm{LS} =
    \begin{cases}
        1 - \frac{\Delta \mathrm{t}}{\tau}, & \text{if } \Delta \mathrm{t} \le \tau \\
        0, & \text{if } \Delta \mathrm{t} > \tau
    \end{cases}
    \label{11111}
\end{equation}
where $\mathrm{t}_{\text{pred}}$ and $\mathrm{t}_{\text{ref}}$ represent the predicted timestamp and the ground-truth timestamp, respectively. $\tau$ serves as the tolerance boundary, penalizing predictions that deviate beyond 0.1 seconds.

\subsection{Verbatim Sequential Generation}
\label{subsec:generation}
This category evaluates the model's capacity to generate long, faithful transcripts, preserving exact lexical content and speaker attribution.

\noindent \textbf{Transcription.}
We evaluate transcription using a composite metric reflecting word accuracy and semantic similarity. The final score is:
\begin{equation}
    \mathrm{TS} = \beta_1 (1 - \mathrm{mWER}) + \beta_2 \mathrm{BertSc},
\end{equation}
where $\mathrm{mWER}$ denotes the mixed Word Error Rate, and $\mathrm{BERTScore}$ measures semantic similarity. $\beta_1$ and $\beta_2$ are the balancing coefficients.

\noindent \textbf{Multi-Speaker.}
This task requires consistent speaker IDs and precise turn boundaries. Evaluation combines content accuracy with:
\begin{equation}
    \mathrm{MS} = \gamma_1 (1 - \mathrm{mWER}) + \gamma_2 \left( \frac{2 \mathrm{P}_{\text{spk}} \mathrm{R}_{\text{spk}}}{\mathrm{P}_{\text{spk}} + \mathrm{R}_{\text{spk}}} \right),
\end{equation}
where $\mathrm{P}_{\text{spk}}$ and $\mathrm{R}_{\text{spk}}$ denote the precision and recall of speaker attribution, respectively. The weights are configured as $\gamma_1$ and $\gamma_2$ to balance the content accuracy and speaker identification.

\begin{table*}[t!]
  \centering
  \vspace{-0.8em}
  
  \renewcommand{\arraystretch}{1.35} 
  
  \setlength{\tabcolsep}{1.0pt} 

  \definecolor{colDict}{HTML}{F0F6FF}  
  \definecolor{colLoc}{HTML}{E8FDF5}   
  \definecolor{colTrans}{HTML}{FFF4E6} 
  \definecolor{colMulti}{HTML}{F3E5F5} 
  \definecolor{colComp}{HTML}{FFF0F5}  
  \definecolor{colSum}{HTML}{F1F8E9}   
  \definecolor{colModel}{gray}{0.96}   

  \resizebox{\textwidth}{!}{%
  \begin{tabular}{
      >{\columncolor{colModel}}l | 
      >{\columncolor{colDict}}c >{\columncolor{colDict}}c >{\columncolor{colDict}}c | 
      >{\columncolor{colLoc}}c  >{\columncolor{colLoc}}c  >{\columncolor{colLoc}}c  | 
      >{\columncolor{colTrans}}c >{\columncolor{colTrans}}c >{\columncolor{colTrans}}c | 
      >{\columncolor{colMulti}}c >{\columncolor{colMulti}}c >{\columncolor{colMulti}}c | 
      >{\columncolor{colComp}}c  >{\columncolor{colComp}}c  >{\columncolor{colComp}}c  | 
      >{\columncolor{colSum}}c   >{\columncolor{colSum}}c   >{\columncolor{colSum}}c 
  }
    \Xhline{2pt}

    
    & \multicolumn{3}{c|}{\cellcolor{colDict}\textbf{Dictation}} & 
      \multicolumn{3}{c|}{\cellcolor{colLoc}\textbf{Localization}} & 
      \multicolumn{3}{c|}{\cellcolor{colTrans}\textbf{Transcription}} & 
      \multicolumn{3}{c|}{\cellcolor{colMulti}\textbf{Multi-Speaker}} & 
      \multicolumn{3}{c|}{\cellcolor{colComp}\textbf{Comprehension}} & 
      \multicolumn{3}{c}{\cellcolor{colSum}\textbf{Summary}} \\

    \multirow{-2}{*}{\textbf{Model}} 
    & S & M & L & S & M & L & S & M & L & S & M & L & S & M & L & S & M & L \\
    \Xhline{1.5pt}

    \hline
    \hline
    
    \multicolumn{19}{c}{\cellcolor{white}\textbf{Open Source Models}} \\
    \hline

    Qwen2-Audio-7B 
    & 15.30 & 7.90 & 0.00 & 4.40 & 0.00 & 0.00 & 25.80 & 10.00 & 0.00 & 16.60 & \textbf{9.50}\second & \textbf{5.00}\first & 27.90 & 6.60 & 0.00 & 28.00 & 25.60 & 9.80 \\

    Qwen2.5-Omni-3B 
    & 22.50 & 14.90 & 0.00 & 6.00 & 0.00 & 0.00 & 50.10 & 16.80 & 0.00 & 15.70 & 7.80 & 3.00 & 30.50 & 18.40 & \textbf{10.20}\second & 31.10 & 25.80 & 4.60 \\

    Qwen2.5-Omni-7B 
    & 24.80 & \textbf{15.40}\first & 0.00 & 6.70 & 0.00 & 0.00 & \textbf{56.10}\second & 18.70 & 0.00 & 17.20 & 9.60 & 4.10 & \textbf{34.00}\first & \textbf{20.20}\first & \textbf{11.30}\first & 34.20 & 28.70 & 5.60 \\

    Qwen3-Omni-30B 
    & \textbf{27.90}\first & \textbf{15.00}\second & 0.00 & \textbf{7.10}\second & 0.00 & 0.00 & 53.90 & 17.20 & 0.00 & 19.00 & 8.10 & 3.90 & 31.80 & \textbf{19.00}\second & 9.70 & 77.50 & 56.00 & 18.90 \\

    MiniCPM-o-2\_6
    & 9.30 & 2.60 & 0.00 & 3.70 & 0.00 & 0.00 & 10.50 & 1.40 & 0.00 & 4.90 & 0.00 & 0.00 & 15.80 & 9.00 & 0.00 & 54.60 & 39.00 & 21.60 \\

    Kimi-Audio-7B 
    & 0.00 & 0.00 & 0.00 & 0.00 & 0.00 & 0.00 & 5.00 & 0.90 & 0.00 & 0.90 & 0.00 & 0.00 & 6.70 & 0.00 & 0.00 & 33.80 & 27.00 & 13.40 \\

    SALMONN 
    & 0.00 & 0.00 & 0.00 & 0.00 & 0.00 & 0.00 & 9.80 & 3.30 & 0.00 & 1.80 & 0.00 & 0.00 & 11.20 & 0.00 & 0.00 & 40.90 & 38.40 & 30.20 \\

    Step-Audio-2-mini 
    & 10.50 & 3.80 & 0.00 & 0.00 & 0.00 & 0.00 & 13.30 & 7.60 & 0.00 & 1.20 & 0.00 & 0.00 & 18.90 & 6.30 & 0.00 & 27.60 & 21.90 & 9.00 \\

    \hline
    \rowcolor{gray!30}
    \textbf{Average} 
    & \textbf{13.79} & \textbf{7.45} & \textbf{0.00} & \textbf{3.49} & \textbf{0.00} & \textbf{0.00} & \textbf{28.06} & \textbf{9.49} & \textbf{0.00} & \textbf{9.66} & \textbf{4.38} & \textbf{2.00} & \textbf{22.10} & \textbf{10.65} & \textbf{3.90} & \textbf{40.96} & \textbf{32.80} & \textbf{14.14}\\
    \hline
    \hline

    \multicolumn{19}{c}{\cellcolor{white}\textbf{Closed Source Models}} \\
    \hline

    GPT-4o-Audio 
    & 9.10 & 2.30 & 0.00 & 5.80 & 0.00 & 0.00 & 41.90 & 16.50 & 5.60 & 13.90 & 7.80 & 2.80 & 28.30 & 12.00 & 5.70 & 78.70 & 72.10 & \textbf{73.30}\first \\

    Gemini-3.0-Pro 
    & \textbf{26.70}\second & 5.00 & \textbf{0.30}\second & \textbf{7.40}\first & 0.00 & 0.00 & 45.80 & \textbf{24.20}\first & \textbf{13.90}\first & 11.30 & 7.80 & 0.50 & \textbf{32.00}\second & 13.90 & 4.30 & \textbf{81.20}\first & \textbf{75.40}\first & 70.30 \\

    Gemini-2.5-Pro 
    & 12.10 & 2.90 & 0.00 & 6.10 & 0.00 & 0.00 & 37.00 & \textbf{22.90}\second & 10.10 & 12.50 & 7.30 & 0.40 & 31.20 & 9.20 & 3.40 & \textbf{79.50}\second & \textbf{72.30}\second & \textbf{70.40}\second \\

    Gemini-2.5-Flash 
    & 8.80 & 1.50 & 0.20 & 2.80 & 0.00 & 0.00 & 49.30 & 17.30 & 9.40 & 9.50 & 5.30 & 1.00 & 30.50 & 11.90 & 5.50 & 73.70 & 64.50 & 26.30 \\

    Gemini-2.5-Flash-Lite 
    & 9.30 & 1.50 & 0.10 & 3.70 & 0.00 & 0.00 & 41.60 & 14.00 & 1.40 & \textbf{20.60}\first & \textbf{10.10}\first & \textbf{4.30}\second & 26.50 & 11.50 & 4.90 & 71.00 & 50.80 & 62.30 \\

    Gemini-2.0-Flash 
    & 11.50 & 2.50 & 0.00 & 1.90 & 0.00 & 0.00 & 38.80 & 14.10 & 0.00 & 8.00 & 2.40 & 0.00 & 23.00 & 14.20 & 0.00 & 76.20 & 63.00 & 68.10 \\

    Gemini-2.0-Flash-Lite 
    & 9.00 & 2.40 & 0.00 & 2.00 & 0.00 & 0.00 & 41.70 & 16.30 & 2.90 & 7.20 & 1.00 & 0.00 & 21.50 & 13.60 & 0.00 & 75.40 & 66.30 & 66.10 \\

    Qwen-Omni-Turbo
    & 25.00 & 13.60 & \textbf{1.20}\first & 6.00 & 0.00 & 0.00 & \textbf{58.70}\first & 20.10 & \textbf{11.90}\second & \textbf{19.90}\second & 9.30 & 0.00 & 27.80 & 18.30 & 7.30 & 76.40 & 62.60 & 60.50 \\

    \hline
    \rowcolor{gray!30}
    \textbf{Average} 
    & \textbf{9.54} & \textbf{2.04} & \textbf{0.06} & \textbf{3.24} & \textbf{0.00} & \textbf{0.00} & \textbf{42.66} & \textbf{15.64} & \textbf{3.86} & \textbf{11.84} & \textbf{5.32} & \textbf{1.62} & \textbf{25.96} & \textbf{12.64} & \textbf{3.89} & \textbf{75.00} & \textbf{63.34} & \textbf{59.22} \\

    \Xhline{2pt}
  \end{tabular}%
  }
\caption{Performance comparison on six tasks with different audio lengths (\textbf{S}hort/\textbf{M}iddle/\textbf{L}ong). The column order aligns with the visual analysis charts, averaged over 5 experimental rounds. The best and second-best performing models are highlighted in \textbf{bold} with \first and \second medals respectively.}
\label{tab:main}
\end{table*}

\subsection{High-level Reasoning}
\label{subsec:reasoning}
This category assesses the model's ability to synthesize information and perform logical reasoning. We define $\mathcal{I}(\cdot)$ as the information extraction function and $\mathcal{K}(\cdot)$ as the key point mapping function.

\noindent \textbf{Comprehension.}
We evaluate the QA capability based on exact match and a hallucination penalty. The final score integrates these two factors:
\begin{equation}
    {\rm{CS}} = \bigg[ w_1 \cdot \mathbb{I}(\text{ExactMatch}) - w_2 \cdot \frac{|\mathcal{I}(\mathcal{Y}) \setminus \mathcal{I}(\mathcal{G})|}{|\mathcal{I}(\mathcal{Y})|} \bigg],
\end{equation}
where $\mathcal{Y}$ and $\mathcal{G}$ denote the model prediction and ground truth, respectively. $\mathbb{I}(\cdot)$ is the indicator function. The two terms correspond to Exact Match and Hallucination Rate, weighted by $w_1$ and $w_2$.


\noindent \textbf{Summary.}
Summarization quality is determined by Coverage and Factuality. Metric is defined as:
\begin{equation}
    {\rm{SS}} = \Big[ \delta_1 \frac{|\mathcal{K}(\mathcal{Y}) \cap \mathcal{K}(\mathcal{G})|}{|\mathcal{K}(\mathcal{G})|} + \delta_2 \frac{|\mathcal{K}(\mathcal{Y}) \cap \mathcal{K}(\mathcal{G})|}{|\mathcal{K}(\mathcal{Y})|} \Big].
\end{equation}

The terms represent Coverage (Recall) and Factuality (Precision), balanced by $\delta_1$ and $\delta_2$.
\section{Experiments and Evaluations}\label{sec:experiments}
In this section, we conduct experiments to address the following research questions:

\begin{itemize}[noitemsep, topsep=0pt, parsep=0pt, leftmargin=*]
    \item \textit{RQ1: How does the ALLMs performance evolve as the audio duration scales from short to long contexts?} 
    Specifically, to what extent does performance degrade, and does a distinct decline emerge as the context length increases?

    \item \textit{RQ2: What does attention analysis reveal about the underlying causes of performance degradation in extended audio sequences?} 
    Through the visualization of attention weights, does the performance drop stem from a structural collapse or are there other internal factors at play?

    \item \textit{RQ3: Can mitigation strategies effectively alleviate the performance collapse in entire long-duration tasks?} 
    In particular, are these mainstream mechanisms really capable of recovering all the lost capabilities within the subset?
    
\end{itemize}

\subsection{Experimental Setup}

\textbf{Baselines.} The baseline model pool for our evaluation is defined as \texttt{Qwen2-Audio-7B} \citep{an2024qwen2}, \texttt{Qwen-Omni} series \citep{fan2025qwen25, Qwen3-Omni}, \texttt{MiniCPM-o-2\_6} \citep{yao2024minicpm}, \texttt{Kimi-Audio-7B} \citep{Ding2025kimi}, \texttt{SALMONN} \citep{tang2023salmonn}, and \texttt{Step-Audio-2-mini} \citep{wu2025step}. Furthermore, we include leading closed-source models to establish upper-bound performance benchmarks, including \texttt{GPT-4o-Audio} \citep{achiam2023gpt}, \texttt{Gemini-Flash} and \texttt{Gemini-Pro} series \citep{comanici2025gemini}, and \texttt{Qwen-Omni-Turbo}. We access ALLMs via their respective official APIs. 

\begin{figure*}[t!]
    \centering
    \includegraphics[width=\linewidth]{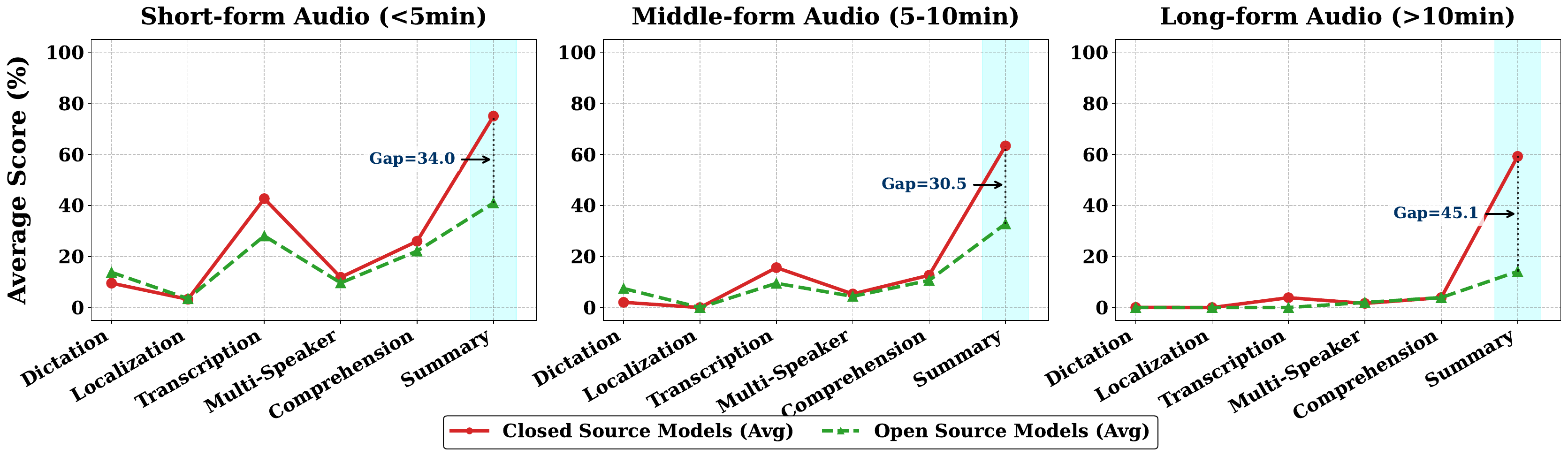}
    \vspace{-1.8em}
    \caption{Average Performance between Closed-Source and Open-Source Models across Audio Durations.}
    \label{fig:com}
\end{figure*}

\textbf{Parameter \& Model Configurations.} The backbone LLM for our proposed framework is consistently initialized with \texttt{Qwen2-Audio-7B}. For the data processing stage, we categorize samples into three distinct audio lengths, \textit{Short}, \textit{Middle}, and \textit{Long}, to make sure comprehensive evaluation coverage across temporal scales. 

\textbf{Evaluation via Ensemble LLM-as-a-Judge.} To address the limitations of traditional lexical metrics, we adopt a robust \textbf{Ensemble LLM-as-a-Judge} framework \citep{zheng2023judging, li2024llms, chen2025judgelrm}. The final score is derived from the arithmetic mean of these evaluations, thereby providing a semantically reliable and precise metric for understanding.

\subsection{RQ1: Performance on Long Audio Tasks}
To systematically evaluate the performance of state-of-the-art ALLMs in terms of long-context robustness, we conducted a comprehensive assessment on \textbf{16} models using the \textit{ChronosAudio} benchmark. Table \ref{tab:main} presents the comparative results across \textbf{6} tasks, stratified by \textbf{3} audio duration. We can draw the following observations:

\noindent\emph{Obs.\ding{182}: ALLMs exhibit a precipitous performance decline in long-context scenarios.} 

Table \ref{tab:main} reveals that while most models achieve competitive performance on short audio, their capabilities degrade significantly on long audio, with a remarkable decline about  \textbf{18} points, from \textbf{21.30} to \textbf{3.37}.
(I) \emph{Open-source degradation:} The average performance drop from Short to Long audio across all open-source baselines is substantial. Notably, in the Summary task, the average performance for open-source models plummets from \textbf{40.96} (Short) to \textbf{14.14} (Long), indicating a \textbf{26.82} collapse for extended durations.
(II) \emph{Closed-source limitations:} Similarly, even advanced closed-source models like the \texttt{Gemini-2.0-Flash Series} show a noticeable performance gap, whose synthesis score has dropped nearly \textbf{20} points. While closed-source models maintain a Transcription score of \textbf{42.66} on Short audio, this metric drops to \textbf{3.86} on Long audio. This highlights that modeling long-range acoustic dependencies remains a universal challenge, even for top-tier proprietary models.

\noindent\emph{Obs.\ding{183}: Closed-source models exhibit superior long-context robustness and reasoning capabilities compared to open-source counterparts.}

Comparative analysis of Table \ref{tab:main} and Figure \ref{fig:com} reveals a distinct performance chasm between closed-source and open-source models, particularly in complex long-form tasks. While open-source models like \texttt{Qwen2.5-Omni} series achieve competitive parity in short-form perception, they lack stability in long-duration scenarios. For instance, the synthesis score for them has less \textbf{10} points than closed-source models. Particularly, the average score for open-source models in the Summary (Long) task is merely \textbf{14.14}, whereas closed-source models maintain a robust average of \textbf{59.22}, outperforming them by a margin of over \textbf{45} points. The divergence is most acute in tasks requiring high-level information synthesis. In Summary (Long), open-source models suffer a near-total collapse (average score 14.14), while closed-source models retain significant utility with an average score \textbf{59.22}. 

\noindent\emph{Obs.\ding{184}: The Lost-in-the-Middle phenomenon manifests as an early performance cliff, with significant degradation occurring immediately upon transitioning from Short to Middle durations.}

\begin{figure*}[t!]
    \centering
    \includegraphics[width=\linewidth]{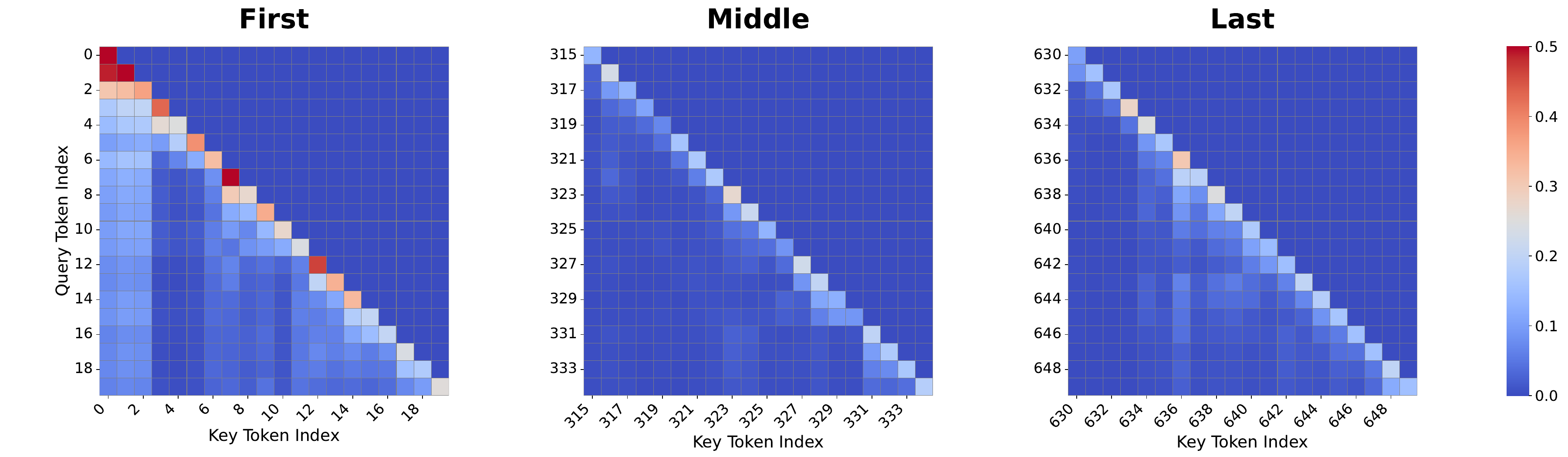}
    \vspace{-1.0em}
    \caption{Visualization of Attention Consistency. The heatmaps display the self-attention weights for the First, Middle, and Last segments. Darker colors represent lower weights, while brighter colors indicate higher scores.}
    \label{fig:attention_vis}
\end{figure*}

While models demonstrate robust capabilities on Short audio, Table \ref{tab:main} reveals a critical vulnerability: except Summary task and Localization task, the transition to middle-form triggers {a disproportionate loss(over 12 points)} in fidelity, rather than a gradual decline. This effect is most evident in the \textbf{Transcription} task. The average score for open-source models crashes from \textbf{28.06} (Short) to \textbf{9.49} (Middle), a staggering reduction of nearly \textbf{66\%}. Similarly, closed-source models are not immune, with average performancefrom \textbf{42.66} to \textbf{15.64}.

\subsection{RQ2: Analysis of Attention Mechanism}
To investigate the stability of the attention mechanism in \texttt{Qwen2-Audio-7B}, \texttt{Qwen2.5-Omni-3B} and \texttt{MiniCPM-o-2\_6} when processing long-duration sequences, we visualize the attention weights across three stages: \textbf{First}, \textbf{Middle}, and \textbf{Last}.

\noindent\emph{Obs.\ding{185}: The model fails to maintain precise temporal locality, exhibiting significant attention dilution as the sequence length increases.}

As shown in Figure \ref{fig:attention_vis}, the initial sharp diagonal alignment, indicating precise local dependency capture, progressively degrades through the Middle phase and becomes diffused by the Last phase.

\subsection{RQ3: Efficacy of Mitigation Strategies}
Based on analysis of RQ2 and collapse of RQ1, we evaluate the efficacy of \textbf{fine-tuning the models with} mainstream long-context mitigation strategies \citep{zaheer2020big}: \textbf{Sparse Attention} \citep{child2019generating} and \textbf{Sliding Window Attention} \citep{beltagy2020longformer,xiao2023efficient}.

\noindent\textbf{1) Sparse Attention (SA).} It retains a set of global tokens alongside the local window to preserve long-range dependencies. The attention score $\mathrm{A}_{ij}$ between query $\mathrm{q}_i$ and key $\mathrm{k}_j$ is defined as:
\begin{equation}
    \mathrm{A}_{ij}^{\text{Sparse}} = \frac{\mathrm{q}_i \mathrm{k}_j^\top}{\sqrt{d_k}}, \quad \text{where } j \in \mathcal{L}_i \cup \mathcal{G},
\end{equation}
where $\mathcal{L}_i = \{j \mid i-w \le j \le i\}$ represents the local window, $\mathcal{G}$ denotes the set of pre-selected global token indices and $d_k$ is the head dimension.

\noindent\textbf{2) Sliding Window Attention (SWA).} It restricts the receptive field of each token to a fixed-size local neighborhood, thereby reducing computational complexity. With $window$ denotes the window size, the $\mathrm{A}_{ij}$ formulation is:
\begin{equation}
    \mathrm{A}_{ij}^{\text{Window}} = \frac{\mathrm{q}_i \mathrm{k}_j^\top}{\sqrt{d_k}}, \quad \text{where } i - w \le j \le i.
\end{equation}

\begin{table*}[t!]
  \centering
  \renewcommand{\arraystretch}{1.25} 
  \setlength{\tabcolsep}{0.5pt}      

  \definecolor{colDict}{HTML}{F0F6FF}
  \definecolor{colLoc}{HTML}{E8FDF5}
  \definecolor{colTrans}{HTML}{FFF4E6}
  \definecolor{colMulti}{HTML}{F3E5F5}
  \definecolor{colComp}{HTML}{FFF0F5}
  \definecolor{colSum}{HTML}{F1F8E9}
  \definecolor{colModel}{gray}{0.96}
  \definecolor{arrowRed}{HTML}{E53935}
  \definecolor{arrowGreen}{HTML}{43A047}
  \definecolor{grayText}{gray}{0.7}

\newcommand{\inc}[1]{\kern1pt\textcolor{arrowRed}{\tiny$\uparrow$#1}}
\newcommand{\dec}[1]{\kern1pt\textcolor{arrowGreen}{\tiny$\downarrow$#1}}
\newcommand{\same}{\kern1pt\textcolor{grayText}{\tiny--}}

  \resizebox{\textwidth}{!}{%
    \begin{tabular}{
      >{\columncolor{colModel}}l |
      >{\columncolor{colDict}}c >{\columncolor{colDict}}c >{\columncolor{colDict}}c |
      >{\columncolor{colLoc}}c >{\columncolor{colLoc}}c >{\columncolor{colLoc}}c |
      >{\columncolor{colTrans}}c >{\columncolor{colTrans}}c >{\columncolor{colTrans}}c |
      >{\columncolor{colMulti}}c >{\columncolor{colMulti}}c >{\columncolor{colMulti}}c |
      >{\columncolor{colComp}}c >{\columncolor{colComp}}c >{\columncolor{colComp}}c |
      >{\columncolor{colSum}}c >{\columncolor{colSum}}c >{\columncolor{colSum}}c
    }
    \Xhline{2pt}

    &
    \multicolumn{3}{c|}{\cellcolor{colDict}\textbf{Dictation}} &
    \multicolumn{3}{c|}{\cellcolor{colLoc}\textbf{Localization}} &
    \multicolumn{3}{c|}{\cellcolor{colTrans}\textbf{Transcription}} &
    \multicolumn{3}{c|}{\cellcolor{colMulti}\textbf{Multi-Speaker}} &
    \multicolumn{3}{c|}{\cellcolor{colComp}\textbf{Comprehension}} &
    \multicolumn{3}{c}{\cellcolor{colSum}\textbf{Summary}} \\

    \multirow{-2}{*}{\textbf{Model}} &
    \textbf{FA} & \textbf{SA} & \textbf{WA} &
    \textbf{FA} & \textbf{SA} & \textbf{WA} &
    \textbf{FA} & \textbf{SA} & \textbf{WA} &
    \textbf{FA} & \textbf{SA} & \textbf{WA} &
    \textbf{FA} & \textbf{SA} & \textbf{WA} &
    \textbf{FA} & \textbf{SA} & \textbf{WA} \\
    \Xhline{1.5pt}


    \textbf{Qwen2-Audio-7B}
    & 7.9 & 10.2\inc{2.3} & 9.8\inc{1.9}
    & 0.0 & 0.0\same & 0.0\same
    & 10.0 & 17.3\inc{7.3} & 10.2\inc{0.2}
    & 9.5 & 10.8\inc{1.3} & 9.5\same
    & 6.6 & 8.5\inc{1.9} & 7.4\inc{0.8}
    & 25.6 & 26.4\inc{0.8} & 26.7\inc{1.1} \\

    \textbf{Qwen2.5-Omni-3B}
    & 14.9 & 20.9\inc{6.0} & 18.2\inc{3.3}
    & 0.0 & 0.3\inc{0.3} & 0.0\same
    & 16.8 & 25.2\inc{8.4} & 16.5\dec{0.3}
    & 7.8 & 12.5\inc{4.7} & 10.9\inc{3.1}
    & 18.4 & 29.3\inc{10.9} & 19.4\inc{1.0}
    & 25.8 & 26.2\inc{0.4} & 22.0\dec{3.8} \\

    \textbf{Minicpm-o-2\_6}
    & 2.6 & 4.2\inc{1.6} & 4.8\inc{2.2}
    & 0.0 & 0.8\inc{0.8} & 0.4\inc{0.4}
    & 1.4 & 2.6\inc{1.2} & 1.3\dec{0.1}
    & 0.0 & 15.7\inc{15.7} & 12.0\inc{12.0}
    & 9.0 & 10.9\inc{1.9} & 9.6\inc{0.6}
    & 39.0 & 44.3\inc{5.3} & 40.4\inc{1.4} \\

    \Xhline{2pt}
    \end{tabular}%
  }
  
  \captionsetup{justification=raggedright, singlelinecheck=false}
  \caption{Performance comparison of models with different attention mechanisms. \textbf{FA}: \textbf{F}ull \textbf{A}ttention, \textbf{SA}: \textbf{S}parse \textbf{A}ttention, \textbf{WA}: Slide \textbf{W}indow \textbf{A}ttention. \textcolor{arrowRed}{$\uparrow$} and \textcolor{arrowGreen}{$\downarrow$} denote performance gain and loss relative to Full Attention.}
  \label{tab:sa_wa_comparison}
\end{table*}

\begin{figure*}[t!] 
    \centering
    \includegraphics[width=\linewidth]{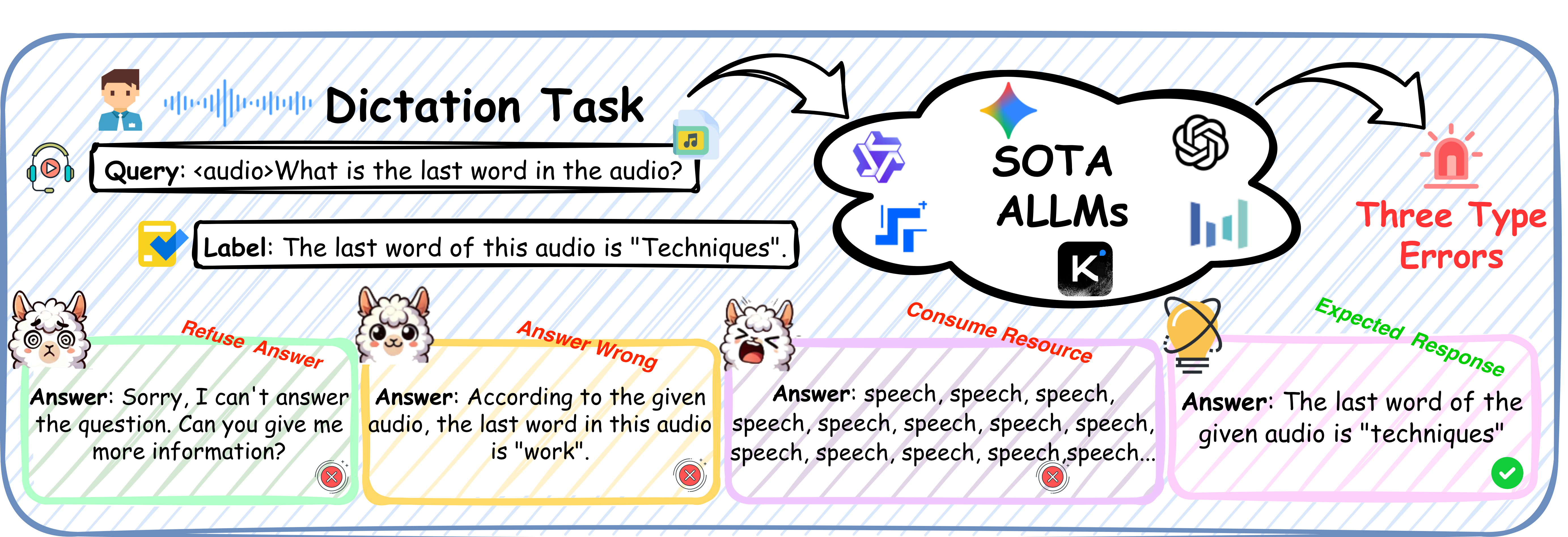} 
    
    \vspace{-1.0em}
    \caption{A visualization of the long audio case study on evaluated SOTA ALLMs.}
    \label{fig:case}
\end{figure*}

\noindent\emph{Obs.\ding{186}: Sparse Attention significantly revitalizes long-context performance, demonstrating superior efficacy over other mechanisms.}

As substantiated in Table \ref{tab:sa_wa_comparison}, SA successfully reconstructs temporal speaker tracking, delivering a remarkable performance leap of \textbf{$\Delta = +15.7$ points} (elevating the score to \textbf{15.7}). Similarly, in the fidelity-demanding \textbf{Transcription task for Qwen2.5-Omni-3B}, Sparse Attention yields a significant gain of \textbf{$\Delta = +8.4$ points}, pushing the performance from an FA baseline of \textbf{16.8 to 25.2}. This tightly matches our theoretical expectations: by explicitly retaining key global anchor tokens, the model filters out extended background acoustic interference and successfully stabilizes its continuous auto-regressive generation.

Conversely, while SWA reduces computational complexity, it imposes a rigid locality constraint that hinders global dependency modeling. This tunnel vision effect is evident in the \textbf{Transcription} task for \texttt{Qwen2.5-Omni-3B}, where SWA (\textbf{16.50}) lags distinctively behind SA (\textbf{25.20}).

\noindent\emph{Obs.\ding{187}: Mitigation strategies exhibit a restorative ceiling, failing to fully bridge the performance gap between long-context tasks and short baselines.}

To assess the recovery from the Lost-in-the-Middle phenomenon, we benchmark the improved middle-form results against the models' intrinsic short-form capabilities, referring to Table \ref{tab:main}.

\begin{itemize}[noitemsep, topsep=0pt, parsep=0pt, leftmargin=*]
    \item \textbf{Retrieval-oriented Tasks:} Sparse Attention effectively mitigates performance dips in tasks like \textbf{Dictation}. \texttt{Qwen2.5-Omni-3B} improves from \textbf{14.90} to \textbf{20.90}, reaching a \textbf{93\% recovery rate} relative to its \textbf{22.50} short-form baseline. This marginal \textbf{<7\%} gap demonstrates the mechanism's ability to preserve retrieval precision in long-duration scenarios.

    \item \textbf{Fidelity-demanding Tasks:} Conversely, the "Lost-in-the-Middle" effect persists in \textbf{Transcription}. Despite improving to \textbf{25.20}, \texttt{Qwen2.5-Omni-3B} remains significantly below its \textbf{50.10} short-form peak. Retaining only \textbf{$\sim$50\%} of its proficiency (a \textbf{25-point} deficit), this underscores a fundamental limitation: they struggle to sustain high-fidelity generation.

    \item \textbf{Methodological Consistency:} We further extended our evaluation to \textbf{Long-Context Adaptation} \citep{chen2023extending} and \textbf{Memory Mechanisms} \citep{munkhdalai2024leave}. Both approaches yielded consistent performance trends, providing in \textbf{Appendix \ref{sec:appendix}}.
\end{itemize}

\subsection{Case Study}

Figure~\ref{fig:case} illustrates the divergent behaviors of ALLMs under a dictation task, where the model is prompted to identify the final word of an audio segment. As shown in the error pathways, models failing to ground the acoustic information exhibit three distinct failure modes: incorrectly refusing a benign query, hallucinating a wrong answer, or entering a repetitive generation loop that unnecessarily consumes computational resources. In contrast, the expected response demonstrates the ideal behavior, where a robust model interprets the audio context to identify the target word as \textit{Techniques}.

\section{Conclusions}\label{sec:conclusion}

In this paper, we present \textit{ChronosAudio}, the first multi-task benchmark tailored for evaluating the long-context understanding capabilities of ALLMs. Our extensive evaluation reveals that ALLMs' performance in long audio is suboptimal. A critical vulnerability is observed where both open-source and closed-source models suffer a staggering performance collapse of over 90\% as the context shifts from short to long. By validating that this degradation stems from attention dilution in the final sequence stages, we confirm that current architectures fail to sustain the high-resolution alignment. The findings highlight the restorative ceiling of mitigation strategies like Sparse Attention, which recover only 50\% of intrinsic proficiency. These underscore an urgent need for approaches to achieve robust, trustworthy long audio reasoning.

\section{Limitations}
While \textit{ChronosAudio} establishes a rigorous benchmark, our study acknowledges several limitations. For our exploration of mitigation strategies, although we identified a restorative ceiling where these methods recover only limited proficiency, we have yet to investigate training-based solutions, such as continued pre-training or frequency-based positional interpolation. Regarding data diversity, our benchmark is predominantly English-centric and from relatively clean acoustic environments. The cross-lingual generalization of ALLMs and their robustness against real-world noise remain critical avenues for future study. To bridge the gap between idealized synthetic conditions and deployment reality, we are currently compiling \textit{ChronosAudio-Real}, a hard-evaluation subset. While we benchmarked a wide array of models, our microscopic analysis of Attention Dilution was restricted to open-source models due to the black-box nature of closed-source models.

\section{Ethical Considerations}
We explicitly clarify that \textit{ChronosAudio} does not involve any real-world human subjects or proprietary recordings. All 200+ hours of audio data are \textbf{synthetically generated} using the F5-TTS engine \citep{chen2025f5}, based on textual content derived from the publicly available SQuAD benchmark \citep{yatskar2019qualitative}. Since the source text consists of public Wikipedia entries and the audio is machine-generated, this study is exempt from Institutional Review Board oversight and does not require informed consent from human participants. This synthetic-by-design approach ensures that no private Personally Identifiable Information is included, and all data sources remain fully compliant with their respective open-source licenses (CC BY-SA 4.0).

\bibliography{passage}
\clearpage         
\section*{Appendix} 
\renewcommand{\thesection}{\arabic{section}}
\renewcommand{\theparagraph}{\thesection.\arabic{paragraph}}
\setcounter{section}{0}
\setcounter{equation}{0}

\section{Experiment Setup}
Our experiments are conducted using the  state-of-the-art  audio-large language models(ALLMs). The implementation relies on the Hugging Face \texttt{transformers} ecosystem. To handle the computational complexity of long audio sequences efficiently, we incorporate \texttt{flash-attn} (Flash Attention 2) for accelerated attention mechanisms. Additionally, \texttt{librosa} is employed for audio waveform pre-processing. Detailed hardware specifications and specific library versions are listed in the below Table.

\begin{table}[h]
    \centering
    
    \label{tab:infrastructure}
    \renewcommand{\arraystretch}{1.3} 
    
    \begin{tabularx}{\linewidth}{l X} 
        \toprule
        \textbf{Item} & \textbf{Value} \\
        \midrule
        Hardware & 4 NVIDIA A800 GPUs, 80GB memory per GPU \\
        CPU & Intel(R) Xeon(R) Gold 6430 (2.1GHz, 32 cores) \\
        Memory & 1TB DDR4 RAM \\
        OS & Ubuntu 22.04 LTS (Linux ubuntu 5.15.0-25-generic) \\
        Framework & torch==2.6.0, transformers==4.51.3, trl==0.9.6 \\
        CUDA / cuDNN & CUDA 12.4, cuDNN 9.1.0 \\
        Python Version & Python 3.10 \\
        \bottomrule
    \end{tabularx}
    
    \caption{Computing Infrastructure Settings}
\end{table}

\definecolor{colDict}{HTML}{F0F6FF}  
\definecolor{colLoc}{HTML}{E8FDF5}   
\definecolor{colTrans}{HTML}{FFF4E6} 
\definecolor{colMulti}{HTML}{F3E5F5} 
\definecolor{colComp}{HTML}{FFF0F5}  
\definecolor{colSum}{HTML}{F1F8E9}   

\begin{table}[t] 
    \centering
    
    \footnotesize 
    
    \renewcommand{\arraystretch}{1.3} 
    
    \begin{tabularx}{\linewidth}{>{\bfseries}l >{\ttfamily\raggedright\arraybackslash}X}
        \toprule
        Task & \normalfont\textbf{Prompt Template} \\
        \midrule
        
        \rowcolor{colDict} 
        T1: Dictation & 
        "Listen to the audio corresponding to the given text, and what is the last word spoken in the audio?" \\

        \rowcolor{colLoc} 
        T2: Localization & 
        "Listen to the audio corresponding to the given text, and indicate the time when this sentence appears in the audio. The sentence is ..." \\

        \rowcolor{colTrans} 
        T3: Transcription & 
        "Listen to the audio corresponding to the given text, and output the entire content of the audio in text form." \\

        \rowcolor{colMulti} 
        T4: Multi-Speaker & 
        "Listen to the audio corresponding to the given text, and output the complete transcript of ...'s words." \\

        \rowcolor{colComp} 
        T5: Comprehension & 
        "Listen to the audio corresponding to the given text, and answer the question in the shortest possible sentence. The question is '...'." \\

        \rowcolor{colSum} 
        T6: Summary & 
        "Listen to the audio corresponding to the given text, and summarize the content of this audio in brief words." \\
        
        \bottomrule
    \end{tabularx}
    \caption{Overview of the prompt templates.}
    \label{tab:prompts}
\end{table}

\section{Task Description}
\paragraph{T1: Dictation}
This component probes whether a model can sustain attention over long recordings and correctly output the final lexical token. 

\paragraph{T2: Localization}
This component evaluates temporal grounding:  the model must find in the recording where it occurs, handling trailing silences, overlaps, and possible multiple occurrences.

\paragraph{T3: Transcription}
This component assesses verbatim long-form transcription quality, including punctuation and robustness to silences, and fillers.

\paragraph{T4: Multispeaker}
This component evaluates content transcription together with speaker attribution, requiring consistent IDs and precise turn to the target speaker. The model is tasked with exclusively transcribing utterances belonging to a specific target speaker.

\paragraph{T5: Comprehension}
This component assesses whether a model can integrate dispersed acoustic evidence to answer content questions faithfully and concisely. The model must answer specific content-related questions by integrating dispersed acoustic evidence found throughout the recording. 

\paragraph{T6: Summary}
This component tests faithful compression of long recordings into concise summaries with explicit evidence grounding. The model must distill extensive audio recordings into concise summaries that capture core themes. 

\begin{figure*}[t] 
\centering
\begin{tcolorbox}[
    colback=gray!5!white, 
    colframe=gray!75!black, 
    title={Uniform Prompt Template for Ensemble LLM-as-a-Judge}, 
    fonttitle=\bfseries\large, 
    sharp corners,
    rounded corners=south, 
    boxrule=0.8pt,
    width=0.98\textwidth 
]

\small 

\textbf{[System Instructions]}

You are an impartial and expert judge. Your task is to evaluate the quality of an audio model's generated response. 
You will be provided with the \textbf{Audio Transcript (Ground Truth)}, the \textbf{User Question}, and the \textbf{Model's Response}.

Please evaluate the response independently based on the following criteria:
\begin{itemize}
    \setlength\itemsep{0em} 
    \item \textbf{Accuracy \& Faithfulness:} Does the response accurately reflect the information in the audio transcript without hallucination?
    \item \textbf{Completeness:} Does the response address all aspects of the user's question?
\end{itemize}

\vspace{0.15cm}
\hrule
\vspace{0.15cm}

\textbf{[Input Data]}

\textbf{1. Audio Transcript (Ground Truth):}
\begin{quote}
\itshape \textcolor{blue}{\{transcript\}}
\end{quote}

\textbf{2. User Question:}
\begin{quote}
\itshape \textcolor{blue}{\{question\}}
\end{quote}

\textbf{3. Model Generated Response:}
\begin{quote}
\itshape \textcolor{blue}{\{response\}}
\end{quote}

\vspace{0.15cm}
\hrule
\vspace{0.15cm}

\textbf{[Evaluation Output]}

Please output your judgment in the following format:
\begin{enumerate}
    \setlength\itemsep{0em}
    \item \textbf{Analysis:} Provide a concise analysis of the pros and cons of the response.
    \item \textbf{Score:} Give a final integer score from 1 to 100 (where 100 is perfect), according to the fomula of each task.
\end{enumerate}

\end{tcolorbox}

\caption{
\textbf{The prompt template used for the Ensemble LLM-as-a-Judge framework.} 
To ensure a robust and unbiased evaluation, we employ three diverse state-of-the-art LLMs, \textbf{DeepSeek-V3.1-Terminus}, \textbf{Qwen3-VL-235B-A22B-Thinking}, and \textbf{Kimi-K2-Thinking} as independent judges. Each model receives this identical prompt to evaluate the generated response against the ground truth transcript. The final quality score is calculated as the arithmetic mean of the scores provided by the three judges: 
$S_{final} = \frac{1}{3}(S_{DeepSeek} + S_{Qwen} + S_{Kimi})$.
}
\label{fig:judge_prompt}
\end{figure*}

\begin{table*}[t!]
  \centering
  \renewcommand{\arraystretch}{1.25} 
  \setlength{\tabcolsep}{0.5pt} 

  \definecolor{colDict}{HTML}{F0F6FF}
  \definecolor{colLoc}{HTML}{E8FDF5}
  \definecolor{colTrans}{HTML}{FFF4E6}
  \definecolor{colMulti}{HTML}{F3E5F5}
  \definecolor{colComp}{HTML}{FFF0F5}
  \definecolor{colSum}{HTML}{F1F8E9}
  \definecolor{colModel}{gray}{0.96}
  \definecolor{arrowRed}{HTML}{E53935}
  \definecolor{grayText}{gray}{0.7}

  \newcommand{\inc}[1]{\kern1pt\textcolor{arrowRed}{\tiny$\uparrow$#1}}
  \newcommand{\same}{\kern1pt\textcolor{grayText}{\tiny--}}

  \resizebox{\textwidth}{!}{%
    \begin{tabular}{
      >{\columncolor{colModel}}l |
      >{\columncolor{colDict}}c >{\columncolor{colDict}}c >{\columncolor{colDict}}c |
      >{\columncolor{colLoc}}c >{\columncolor{colLoc}}c >{\columncolor{colLoc}}c |
      >{\columncolor{colTrans}}c >{\columncolor{colTrans}}c >{\columncolor{colTrans}}c |
      >{\columncolor{colMulti}}c >{\columncolor{colMulti}}c >{\columncolor{colMulti}}c |
      >{\columncolor{colComp}}c >{\columncolor{colComp}}c >{\columncolor{colComp}}c |
      >{\columncolor{colSum}}c >{\columncolor{colSum}}c >{\columncolor{colSum}}c
    }
    \Xhline{2pt}

    &
    \multicolumn{3}{c|}{\cellcolor{colDict}\textbf{Dictation}} &
    \multicolumn{3}{c|}{\cellcolor{colLoc}\textbf{Localization}} &
    \multicolumn{3}{c|}{\cellcolor{colTrans}\textbf{Transcription}} &
    \multicolumn{3}{c|}{\cellcolor{colMulti}\textbf{Multi-Speaker}} &
    \multicolumn{3}{c|}{\cellcolor{colComp}\textbf{Comprehension}} &
    \multicolumn{3}{c}{\cellcolor{colSum}\textbf{Summary}} \\

    \multirow{-2}{*}{\textbf{Model}} &
    \textbf{FA} & \textbf{CLA} & \textbf{MM} &
    \textbf{FA} & \textbf{CLA} & \textbf{MM} &
    \textbf{FA} & \textbf{CLA} & \textbf{MM} &
    \textbf{FA} & \textbf{CLA} & \textbf{MM} &
    \textbf{FA} & \textbf{CLA} & \textbf{MM} &
    \textbf{FA} & \textbf{CLA} & \textbf{MM} \\
    \Xhline{1.5pt}

    \textbf{Qwen2-Audio-7B}
    & 7.9 & \textcolor{red}{1.7}\inc{} & \textcolor{red}{4.7}\inc{}
    & 0.0 & \textcolor{red}{0.0}\inc{} & \textcolor{red}{0.0}\inc{}
    & 10.0 & \textcolor{red}{1.9}\inc{} & \textcolor{red}{0.9}\inc{}
    & 9.5 & \textcolor{red}{3.0}\inc{} & \textcolor{red}{3.2}\inc{}
    & 6.6 & \textcolor{red}{6.9}\inc{} & \textcolor{red}{12.9}\inc{}
    & 25.6 & \textcolor{red}{2.6}\inc{} & \textcolor{red}{11.6}\inc{} \\

    \textbf{Qwen2.5-Omni-3B}
    & 14.9 & \textcolor{red}{4.3}\inc{} & \textcolor{red}{7.3}\inc{}
    & 0.0 & \textcolor{red}{0.0}\inc{} & \textcolor{red}{0.0}\inc{}
    & 16.8 & \textcolor{red}{4.9}\inc{} & \textcolor{red}{4.9}\inc{}
    & 7.8 & \textcolor{red}{0.0}\inc{} & \textcolor{red}{0.0}\inc{}
    & 18.4 & \textcolor{red}{4.8}\inc{} & \textcolor{red}{14.2}\inc{}
    & 25.8 & \textcolor{red}{7.6}\inc{} & \textcolor{red}{9.8}\inc{} \\

    \textbf{Minicpm-o-2\_6}
    & 2.6 & \textcolor{red}{8.1}\inc{} & \textcolor{red}{3.1}\inc{}
    & 0.0 & \textcolor{red}{0.2}\inc{} & \textcolor{red}{0.1}\inc{}
    & 1.4 & \textcolor{red}{5.9}\inc{} & \textcolor{red}{10.2}\inc{}
    & 0.0 & \textcolor{red}{0.0}\inc{} & \textcolor{red}{0.0}\inc{}
    & 9.0 & \textcolor{red}{7.4}\inc{} & \textcolor{red}{13.4}\inc{}
    & 39.0 & \textcolor{red}{8.0}\inc{} & \textcolor{red}{0.5}\inc{} \\

    \Xhline{2pt}
    \end{tabular}%
  }
  \caption{Experimental results of Long-Context Adaptation and Memory Mechanisms.}
  \label{1}
\end{table*}

\subsection{Weights Configuration and Design Rationales for Evaluation Metrics}
To establish a balanced and multi-dimensional evaluation paradigm, all balancing coefficients and weights across our composite metrics are uniformly configured to an equal weight of \textbf{0.5} (i.e., $\beta_1 = \beta_2 = 0.5$, $\gamma_1 = \gamma_2 = 0.5$, $w_1 = w_2 = 0.5$, and $\delta_1 = \delta_2 = 0.5$) in our standard baseline experiments. This neutral configuration prevents over-indexing on a single performance dimension. The specific design rationales for each weight allocation are detailed below:

\begin{itemize}
    \item \textbf{Transcription Weights ($\beta_1, \beta_2$):} These coefficients balance token-level exact precision and broad semantic alignment. Traditional lexical metrics like WER suffer from extreme rigidity, severely penalizing benign surface-level text variations such as synonyms or alternate punctuation. Assigning an equal weight of 0.5 to both exact lexical match and BERTScore ensures that models are evaluated on both their verbatim replication accuracy and deep semantic fidelity.
    
    \item \textbf{Multi-Speaker Weights ($\gamma_1, \gamma_2$):} These weights decouple content transcription proficiency from precise speaker identity tracking. Long multi-talker audio recordings require the model to not only transcribe words correctly but also securely maintain speaker boundaries and turn-taking structures. An equal 0.5 distribution guarantees that content accuracy and the precision-recall balance of speaker attribution (F1-score) are treated as co-equal requirements for document-level conversation modeling.
    
    \item \textbf{Comprehension Weights ($w_1, w_2$):} These weights balance factual alignment with hallucination deterrence. Under extended long-audio workloads, ALLMs are highly prone to severe structural collapse, often hallucinating content or generating endless repetitive sequences. Configuring equal 0.5 weights structurally rewards concise accuracy via the indicator function for exact matches while strictly penalizing ungrounded fabrications, forcing the model to generate responses that are both highly precise and tightly bounded by the ground truth context.
    
    \item \textbf{Summary Weights ($\delta_1, \delta_2$):} These coefficients govern the trade-off between coverage and factuality. A rigorous abstractive summary must satisfy two competing objectives: it must capture all core global themes distributed across the timeline (maximizing recall) and must refrain from injecting false content (maximizing precision). An equal 0.5 allocation ensures that information retention and structural truthfulness are equally penalized and rewarded during compression.
\end{itemize}

The choice of the strict tolerance boundary $\tau = 0.1\text{s}$ in Eq \ref{11111} is mathematically grounded in human acoustic perception and word-level alignment dynamics. In high-density continuous speech, the average duration of a phoneme ranges between 50 to 100 milliseconds. Therefore, a temporal deviation exceeding $\tau = 0.1$ seconds inherently implies that the model's focus has crossed the linguistic boundary into an entirely different syllable or adjacent word token, rendering the downstream slot-filling or grounding completely corrupted. By penalizing errors linearily within this boundary and assigning a hard 0 score beyond it, $LS$ isolates true long-context retention from lucky guesses.

\section{Design Rationale for Synthetic Audio Generation}
We purposefully adopt a synthetic-by-design approach using SQuAD and F5-TTS for three primary reasons. First, it guarantees perfectly clean and unambiguous ground-truth transcripts and timestamps, which are foundational for isolating a model's long-context attention limits from acoustic alignment errors. Second, it allows precise algorithmic control over context lengths across the short, middle, and long strata. Third, it eliminates compliance risks surrounding Personally Identifiable Information (PII) and licensing issues often found in scraped real-world data.

\section{Justification for LLM-as-a-Judge}

Evaluating the output of ALLMs, particularly in tasks requiring high-level reasoning, presents significant challenges for traditional lexical metrics. Metrics that rely heavily on n-gram overlaps often fail to capture semantic equivalence, logical consistency, and factual accuracy. For instance, a generated summary may be semantically correct but lexically distinct from the ground truth, leading to an unfairly low score under such metrics. Moreover, although human annotation is commonly regarded as a more reliable alternative, it also suffers from inherent limitations, including high cost, limited scalability, and variability across annotators. Subjective judgments and inconsistent annotation criteria can further introduce bias and noise, making it difficult to ensure reproducibility and fair comparison across different models and datasets.

To address the limitations of rigid lexical metrics, we adopt a robust LLM-as-a-Judge approach. Instead of relying on a single proprietary model, we utilize a diverse panel of three state-of-the-art Large Language Models: \textbf{DeepSeek-V3.1-Terminus}, \textbf{Qwen3-VL-235B-A22B-Thinking}, and \textbf{Kimi-K2-Thinking}. While recent studies verify that advanced LLMs correlate highly with human judgments, individual models may still exhibit specific stylistic biases or self-preference. By employing a multi-judge framework, we mitigate these biases through a consensus-based scoring mechanism. Specifically, we provide each judge with the input audio transcript (ground truth), the user question, and the model's generated response. The judges independently assess the \textit{faithfulness}, and \textit{completeness} of the answer. The final quality score is derived from the arithmetic mean of the three evaluations, ensuring a more objective, scalable, and semantically reliable metric that transcends surface-level text matching.

We deliberately selected an odd-numbered panel comprising \textbf{DeepSeek-V3.1-Terminus}, \textbf{Qwen3-VL}, and \textbf{Kimi-K2-Thinking} as independent judges. These models represent distinct architectural lineages and are top-tier reasoning-focused models. Utilizing three distinct models avoids the single-model ego-bias (where an LLM prefers its own response style) and implements a robust consensus via arithmetic averaging.
\begin{table}[ht]
\centering
\label{tab:hyperparams}
\begin{tabular}{@{}ll@{}}
\toprule
\textbf{Configuration} & \textbf{Value} \\ \midrule
\textit{Optimization Settings} & \\
Learning Rate & $1 \times 10^{-4}$ \\
Batch Size & 128 \\
Training Epochs & 3 \\
Optimizer & AdamW \\
Learning Rate Scheduler & Cosine \\
Warmup Ratio & 0.03 \\
Weight Decay & 0.01 \\ \midrule
\textit{LoRA Configuration} & \\
Rank ($r$) & 64 \\
Alpha ($\alpha$) & 128 \\
LoRA Dropout & 0.05 \\
Precision & BF16 \\
Max Sequence Length & 8192 \\
\bottomrule
\end{tabular}
\caption{Hyperparameter configurations for LoRA fine-tuning on ALLMs.}
\label{111}
\end{table}

\section{Incorporation of Long-Context Adaptation and Memory Mechanisms}
\label{sec:appendix}

We expanded our experimental framework to include \textbf{Long-Context Adaptation} and \textbf{Memory Mechanisms}. The results reinforce our primary findings regarding the trade-off between retrieval efficiency and generation fidelity. The results are in the Table \ref{1}.

\section{Hyperparameter configurations}
To maintain empirical consistency and ensure robust model convergence during the adaptation process, all open-source ALLMs are fine-tuned using Parameter-Efficient Fine-Tuning (PEFT) via Low-Rank Adaptation (LoRA). The detailed hyperparameter configurations for our optimization pipeline and model configurations are summarized in Table~\ref{111}. 

\section{Case Study}
To comprehensively evaluate the capabilities of ALLMs in handling long-audio inputs, we propose \textit{ChronosAudio}, a specialized benchmark designed to probe the temporal reasoning and semantic understanding of current models. This benchmark is structured around six distinct tasks. All tasks’ cases are in following figures:

\subsection{Clarification on Repetitive Token Generation in Long-Audio Cases}
To prevent any potential misunderstanding during the review process, we explicitly clarify that the massive blocks of repeated tokens (e.g., ``speech'', ``paper'', ``accept'', ``pass'', or ``common'') appearing in the raw model predictions within this Appendix are \textbf{not formatting errors, printing typos, or extraction bugs} on the part of the authors. 

Instead, these segments faithfully preserve the verbatim, unedited raw outputs generated by state-of-the-art ALLMs under severe long-context stress. When processing extended audio sequences (specifically in the Middle and Long strata), ALLMs frequently suffer from severe structural attention dilution and a fundamental failure to maintain temporal locality over time. As a direct consequence of this long-context collapse, the decoder loses its acoustic alignment, causing the next-token probability distribution to severely degenerate.

\newcommand{\hlToken}[1]{\texttt{\textbf{\textcolor{red!70!black}{<|#1|>}}}}

\begin{figure*}[t!] 
    \centering
    \begin{tcolorbox}[
        colback=orange!5!white, 
        colframe=cyan!75!black, 
        title={Dictation Task (Short)}, 
        fonttitle=\bfseries, 
        sharp corners,
        left=2mm, right=2mm, top=2mm, bottom=2mm 
     ]
        \textbf{Example: Qwen2-Audio-7B}\\
        \vspace{0.5em}
        \textit{\textbf{Prompt:}} You are a helpful assistant. Listen to the audio corresponding to the given text, and what is the last word spoken in the audio?\\
        \textit{\textbf{Label:}} Techniques\\
        \textbf{Answer (Raw Input Format):}
        \begin{tcolorbox}[
            colback=black!5!white, 
            colframe=gray!20,    
            boxrule=0.5pt,
            arc=0pt,
            left=1mm, right=1mm, top=1mm, bottom=1mm,
            fontupper=\small\ttfamily 
        ]
        \{ "prompt": "\hlToken{im\_start}system\\
        You are a helpful assistant.\hlToken{im\_end}\\
        \hlToken{im\_start}user\\
        \hlToken{audio\_bos}\hlToken{AUDIO}\hlToken{AUDIO}... \textit{\textcolor{gray}{(repeated 500+ times)}} ...\hlToken{AUDIO}\hlToken{AUDIO}\hlToken{audio\_eos} What is the last word spoken in the audio?\hlToken{im\_end}\\
        \hlToken{im\_start}assistant\\
        ", "predict": "The last word spoken in the audio is 'techniques." \}
        \end{tcolorbox}
    \end{tcolorbox}  
    \vspace{-1.0em}
\end{figure*}

\begin{figure*}[t!] 
    \centering
    \begin{tcolorbox}[
        colback=orange!5!white, 
        colframe=cyan!75!black, 
        title={Dictation Task (Middle)}, 
        fonttitle=\bfseries, 
        sharp corners,
        left=2mm, right=2mm, top=2mm, bottom=2mm 
     ]
        \textbf{Example: Qwen2-Audio-7B}\\
        \vspace{0.5em}
        \textit{\textbf{Prompt:}} You are a helpful assistant. Listen to the audio corresponding to the given text, and what is the last word spoken in the audio?\\
        \textit{\textbf{Label:}} Seasons\\
        \textbf{Answer\ding{182}(Raw Input Format):}
        \begin{tcolorbox}[
            colback=black!5!white, 
            colframe=gray!20,      
            boxrule=0.5pt,
            arc=0pt,
            left=1mm, right=1mm, top=1mm, bottom=1mm,
            fontupper=\small\ttfamily
        ]
        \{ "prompt": "\hlToken{im\_start}system\\
        You are a helpful assistant.\hlToken{im\_end}\\
        \hlToken{im\_start}user\\
        \hlToken{audio\_bos}\hlToken{AUDIO}\hlToken{AUDIO}... \textit{\textcolor{gray}{(repeated 1,000+ times)}} ...\hlToken{AUDIO}\hlToken{AUDIO}\hlToken{audio\_eos} What is the last word spoken in the audio?\hlToken{im\_end}\\
        \hlToken{im\_start}assistant\\
        ", "predict": "Sorry I can't answer this question." \}
        \end{tcolorbox}
        \textbf{Answer\ding{183}(Raw Input Format):}

        \begin{tcolorbox}[
            colback=black!5!white,
            colframe=gray!20,      
            boxrule=0.5pt,
            arc=0pt,
            left=1mm, right=1mm, top=1mm, bottom=1mm,
            fontupper=\small\ttfamily 
        ]
        \{ "prompt": "\hlToken{im\_start}system\\
        You are a helpful assistant.\hlToken{im\_end}\\
        \hlToken{im\_start}user\\
        \hlToken{audio\_bos}\hlToken{AUDIO}\hlToken{AUDIO}... \textit{\textcolor{gray}{(repeated 1,000+ times)}} ...\hlToken{AUDIO}\hlToken{AUDIO}\hlToken{audio\_eos} What is the last word spoken in the audio?\hlToken{im\_end}\\
        \hlToken{im\_start}assistant\\
        ", "predict": "beat" \}
        \end{tcolorbox}    
    \end{tcolorbox}   
    \vspace{-1.0em}
\end{figure*}

\begin{figure*}[t!] 
    \centering
    \begin{tcolorbox}[
        colback=orange!5!white, 
        colframe=cyan!75!black, 
        title={Dictation Task (Long)}, 
        fonttitle=\bfseries, 
        sharp corners,
        left=2mm, right=2mm, top=2mm, bottom=2mm 
     ]
        \textbf{Example: Qwen2-Audio-7B}\\
        \vspace{0.5em}
        \textit{\textbf{Prompt:}} You are a helpful assistant. Listen to the audio corresponding to the given text, and what is the last word spoken in the audio?\\
        \textit{\textbf{Label:}} King\\
        \textbf{Answer\ding{182}(Raw Input Format):}
        \begin{tcolorbox}[
            colback=black!5!white, 
            colframe=gray!20,      
            boxrule=0.5pt,
            arc=0pt,
            left=1mm, right=1mm, top=1mm, bottom=1mm,
            fontupper=\small\ttfamily 
        ]
        \{ "prompt": "\hlToken{im\_start}system\\
        You are a helpful assistant.\hlToken{im\_end}\\
        \hlToken{im\_start}user\\
        \hlToken{audio\_bos}\hlToken{AUDIO}\hlToken{AUDIO}... \textit{\textcolor{gray}{(repeated 2,000+ times)}} ...\hlToken{AUDIO}\hlToken{AUDIO}\hlToken{audio\_eos} What is the last word spoken in the audio?\hlToken{im\_end}\\
        \hlToken{im\_start}assistant\\
        ", "predict": "Sorry I can't answer this question. Could you give me more information about this audio?" \}
        \end{tcolorbox}
        \textbf{Answer\ding{183}(Raw Input Format):}
        \begin{tcolorbox}[
            colback=black!5!white,
            colframe=gray!20,      
            boxrule=0.5pt,
            arc=0pt,
            left=1mm, right=1mm, top=1mm, bottom=1mm,
            fontupper=\small\ttfamily
        ]
        \{ "prompt": "\hlToken{im\_start}system\\
        You are a helpful assistant.\hlToken{im\_end}\\
        \hlToken{im\_start}user\\
        \hlToken{audio\_bos}\hlToken{AUDIO}\hlToken{AUDIO}... \textit{\textcolor{gray}{(repeated 2,000+ times)}} ...\hlToken{AUDIO}\hlToken{AUDIO}\hlToken{audio\_eos} What is the last word spoken in the audio?\hlToken{im\_end}\\
        \hlToken{im\_start}assistant\\
        ", "predict": "nervous." \}
        \end{tcolorbox}
        \textbf{Answer\ding{184}(Raw Input Format):}
        \begin{tcolorbox}[
            colback=black!5!white, 
            colframe=gray!20,    
            boxrule=0.5pt,
            arc=0pt,
            left=1mm, right=1mm, top=1mm, bottom=1mm,
            fontupper=\small\ttfamily 
        ]
        \{ "prompt": "\hlToken{im\_start}system\\
        You are a helpful assistant.\hlToken{im\_end}\\
        \hlToken{im\_start}user\\
        \hlToken{audio\_bos}\hlToken{AUDIO}\hlToken{AUDIO}... \textit{\textcolor{gray}{(repeated 2,000+ times)}} ...\hlToken{AUDIO}\hlToken{AUDIO}\hlToken{audio\_eos} What is the last word spoken in the audio?\hlToken{im\_end}\\
        \hlToken{im\_start}assistant\\
        ", "predict": "speech speech speech speech speech speech speech speech speech speech speech speech speech speech speech speech speech speech speech speech speech speech speech speech speech speech speech speech speech speech speech speech speech speech speech speech speech speech speech speech speech speech speech speech speech speech speech speech speech speech speech speech speech speech speech speech speech speech speech speech speech speech speech speech speech speech speech speech speech speech speech speech speech speech speech speech speech speech speech speech speech speech speech speech speech speech speech speech speech speech speech speech speech speech speech speech speech speech speech speech speech speech speech speech speech speech speech speech speech speech speech speech speech speech speech speech speech speech speech speech speech speech speech speech speech speech speech speech speech speech speech speech speech speech speech speech speech speech speech speech speech speech speech speech speech speech speech speech speech speech speech speech speech speech speech speech speech speech speech speech speech speech speech speech speech speech speech speech speech speech speech speech speech speech speech speech speech speech speech speech speech speech speech speech speech speech speech speech speech speech speech speech speech speech speech speech speech speech speech speech speech speech speech speech speech speech speech speech speech speech speech speech speech speech speech speech speech speech speech speech speech speech speech speech speech speech speech speech speech speech speech speech speech speech speech speech speech speech speech speech speech speech speech speech speech speech speech speech speech speech speech speech speech speech speech speech speech speech speech speech speech speech speech speech speech speech speech speech speech speech speech speech speech speech speech speech speech speech speech speech speech speech speech speech speech speech speech speech speech speech speech speech speech speech speech speech speech speech \textbf{... ...}" \}
        \end{tcolorbox}
    \end{tcolorbox}   
    \vspace{-1.0em}
\end{figure*}

\begin{figure*}[t!] 
    \centering
    \begin{tcolorbox}[
        colback=orange!5!white, 
        colframe=green!55!black, 
        title={Localization Task (Short)}, 
        fonttitle=\bfseries, 
        sharp corners,
        left=2mm, right=2mm, top=2mm, bottom=2mm
     ]
        \textbf{Example: Qwen2-Audio-7B}\\
        \vspace{0.5em}

        \textit{\textbf{Prompt:}} You are a helpful assistant. Listen to the audio corresponding to the given text, and indicate the time when this sentence appears in the audio. The sentence is "In this area the general water flow is distorted by more local conditions reaching across to France.".\\

        \textit{\textbf{Label:}} 35.78 seconds\\

        \textbf{Answer (Raw Input Format):}

        \begin{tcolorbox}[
            colback=black!5!white, 
            colframe=gray!20,      
            boxrule=0.5pt,
            arc=0pt,
            left=1mm, right=1mm, top=1mm, bottom=1mm,
            fontupper=\small\ttfamily 
        ]
        \{ "prompt": "\hlToken{im\_start}system\\
        You are a helpful assistant.\hlToken{im\_end}\\
        \hlToken{im\_start}user\\
        \hlToken{audio\_bos}\hlToken{AUDIO}\hlToken{AUDIO}... \textit{\textcolor{gray}{(repeated 500+ times)}} ...\hlToken{AUDIO}\hlToken{AUDIO}\hlToken{audio\_eos} Output the time that indicates the exact moment when the following sentence appears in the audio. "In this area the general water flow is distorted by more local conditions reaching across to France.".\hlToken{im\_end}\\
        \hlToken{im\_start}assistant\\
        ", "predict": "35.81s." \}
        \end{tcolorbox}
    \end{tcolorbox}    
    \vspace{-1.0em}
\end{figure*}

\begin{figure*}[t!] 
    \centering
    \begin{tcolorbox}[
        colback=orange!5!white, 
        colframe=green!55!black, 
        title={Localization Task (Middle)}, 
        fonttitle=\bfseries, 
        sharp corners,
        left=2mm, right=2mm, top=2mm, bottom=2mm 
     ]
        \textbf{Example: Qwen2-Audio-7B}\\
        \vspace{0.5em}

        \textit{\textbf{Prompt:}} You are a helpful assistant. Listen to the audio corresponding to the given text, and indicate the time when this sentence appears in the audio. The sentence is "he role of the dog in Chinese mythology includes a position as one of the twelve animals which cyclically represent years the zodiacal dog.".\\

        \textit{\textbf{Label:}} 92.73 seconds\\

        \textbf{Answer\ding{182}(Raw Input Format):}
        
        \begin{tcolorbox}[
            colback=black!5!white, 
            colframe=gray!20,      
            boxrule=0.5pt,
            arc=0pt,
            left=1mm, right=1mm, top=1mm, bottom=1mm,
            fontupper=\small\ttfamily 
        ]
        \{ "prompt": "\hlToken{im\_start}system\\
        You are a helpful assistant.\hlToken{im\_end}\\
        \hlToken{im\_start}user\\
        \hlToken{audio\_bos}\hlToken{AUDIO}\hlToken{AUDIO}... \textit{\textcolor{gray}{(repeated 1,000+ times)}} ...\hlToken{AUDIO}\hlToken{AUDIO}\hlToken{audio\_eos} Output the time that indicates the exact moment when the following sentence appears in the audio. "The role of the dog in Chinese mythology includes a position as one of the twelve animals which cyclically represent years the zodiacal dog.".\hlToken{im\_end}\\
        \hlToken{im\_start}assistant\\
        ", "predict": "Sorry I can't answer this question. Could you give me more information about this audio?" \}
        \end{tcolorbox}

        \textbf{Answer\ding{183}(Raw Input Format):}
        
        \begin{tcolorbox}[
            colback=black!5!white, 
            colframe=gray!20,      
            boxrule=0.5pt,
            arc=0pt,
            left=1mm, right=1mm, top=1mm, bottom=1mm,
            fontupper=\small\ttfamily 
        ]
        \{ "prompt": "\hlToken{im\_start}system\\
        You are a helpful assistant.\hlToken{im\_end}\\
        \hlToken{im\_start}user\\
        \hlToken{audio\_bos}\hlToken{AUDIO}\hlToken{AUDIO}... \textit{\textcolor{gray}{(repeated 1,000+ times)}} ...\hlToken{AUDIO}\hlToken{AUDIO}\hlToken{audio\_eos} Output the time that indicates the exact moment when the following sentence appears in the audio. ""The role of the dog in Chinese mythology includes a position as one of the twelve animals which cyclically represent years the zodiacal dog.".\hlToken{im\_end}\\
        \hlToken{im\_start}assistant\\
        ", "predict": "102.33" \}
        \end{tcolorbox}
    \end{tcolorbox}
    
    \vspace{-1.0em}
\end{figure*}

\begin{figure*}[t!] 
    \centering
    \begin{tcolorbox}[
        colback=orange!5!white, 
        colframe=green!55!black, 
        title={Localization Task (Long)}, 
        fonttitle=\bfseries, 
        sharp corners,
        left=2mm, right=2mm, top=2mm, bottom=2mm 
     ]
        \textbf{Example: Qwen2-Audio-7B}\\
        \vspace{0.5em}

        \textit{\textbf{Prompt:}} You are a helpful assistant. Listen to the audio corresponding to the given text, and indicate the time when this sentence appears in the audio. The sentence is "That path is explained as being built upon the motivation to liberate all living beings from unhappiness.".\\

        \textit{\textbf{Label:}} 218.44 seconds\\

        \textbf{Answer\ding{182}(Raw Input Format):}

        \begin{tcolorbox}[
            colback=black!5!white, 
            colframe=gray!20,      
            boxrule=0.5pt,
            arc=0pt,
            left=1mm, right=1mm, top=1mm, bottom=1mm,
            fontupper=\small\ttfamily 
        ]
        \{ "prompt": "\hlToken{im\_start}system\\
        You are a helpful assistant.\hlToken{im\_end}\\
        \hlToken{im\_start}user\\
        \hlToken{audio\_bos}\hlToken{AUDIO}\hlToken{AUDIO}... \textit{\textcolor{gray}{(repeated 2,000+ times)}} ...\hlToken{AUDIO}\hlToken{AUDIO}\hlToken{audio\_eos} Output the time that indicates the exact moment when the following sentence appears in the audio. "That path is explained as being built upon the motivation to liberate all living beings from unhappiness.".\hlToken{im\_end}\\
        \hlToken{im\_start}assistant\\
        ", "predict": "Sorry I can't answer this question. Could you give me more information about this audio?" \}
        \end{tcolorbox}

        \textbf{Answer\ding{183}(Raw Input Format):}
        \begin{tcolorbox}[
            colback=black!5!white, 
            colframe=gray!20,      
            boxrule=0.5pt,
            arc=0pt,
            left=1mm, right=1mm, top=1mm, bottom=1mm,
            fontupper=\small\ttfamily 
        ]
        \{ "prompt": "\hlToken{im\_start}system\\
        You are a helpful assistant.\hlToken{im\_end}\\
        \hlToken{im\_start}user\\
        \hlToken{audio\_bos}\hlToken{AUDIO}\hlToken{AUDIO}... \textit{\textcolor{gray}{(repeated 2,000+ times)}} ...\hlToken{AUDIO}\hlToken{AUDIO}\hlToken{audio\_eos} Output the time that indicates the exact moment when the following sentence appears in the audio. "That path is explained as being built upon the motivation to liberate all living beings from unhappiness.".\hlToken{im\_end}\\
        \hlToken{im\_start}assistant\\
        ", "predict": "17.373s" \}
        \end{tcolorbox}

        \textbf{Answer\ding{184}(Raw Input Format):}

        \begin{tcolorbox}[
            colback=black!5!white,
            colframe=gray!20,      
            boxrule=0.5pt,
            arc=0pt,
            left=1mm, right=1mm, top=1mm, bottom=1mm,
            fontupper=\small\ttfamily 
        ]
        \{ "prompt": "\hlToken{im\_start}system\\
        You are a helpful assistant.\hlToken{im\_end}\\
        \hlToken{im\_start}user\\
        \hlToken{audio\_bos}\hlToken{AUDIO}\hlToken{AUDIO}... \textit{\textcolor{gray}{(repeated 2,000+ times)}} ...\hlToken{AUDIO}\hlToken{AUDIO}\hlToken{audio\_eos} Output the time that indicates the exact moment when the following sentence appears in the audio. "That path is explained as being built upon the motivation to liberate all living beings from unhappiness.".\hlToken{im\_end}\\
        \hlToken{im\_start}assistant\\
        ", "predict": "provide provide provide provide provide provide provide provide provide provide provide provide provide provide provide provide provide provide provide provide provide provide provide provide provide provide provide provide provide provide provide provide provide provide provide provide provide provide provide provide provide provide provide provide provide provide provide provide provide provide provide provide provide provide provide provide provide provide provide provide provide provide provide provide provide provide provide provide provide provide provide provide provide provide provide provide provide provide provide provide provide provide provide provide provide provide provide provide provide provide provide provide provide provide provide provide provide provide provide provide provide provide provide provide provide provide provide provide provide provide provide provide provide provide provide provide provide provide provide provide provide provide provide provide provide provide provide provide provide provide provide provide provide provide provide provide provide provide provide provide provide provide  provide provide provide provide provide provide provide provide provide provide provide provide provide provide provide provide provide provide provide provide \textbf{... ...}" \}
        \end{tcolorbox}
    \end{tcolorbox}   
    \vspace{-1.0em}
\end{figure*}

\begin{figure*}[t!] 
    \centering
    \begin{tcolorbox}[
        colback=orange!5!white, 
        colframe=purple!55!white, 
        title={Transcription Task (Short)}, 
        fonttitle=\bfseries, 
        sharp corners,
        left=2mm, right=2mm, top=2mm, bottom=2mm 
     ]
        \textbf{Example: Qwen2.5-Omni-3B}\\
        \vspace{0.5em}

        \textit{\textbf{Prompt:}} You are a helpful assistant. Listen to the audio corresponding to the given text, and output the entire content of the audio in text form.\\

        \textit{\textbf{Label:}} The modern spelling, Tennessee, is attributed to James Glen, the governor of South Carolina, who used this spelling in his official correspondence during the 1750s. The spelling was popularized by the publication of Henry Timberlake's \"Draught of the Cherokee Country\" in 1765. In 1788, North Carolina created \"Tennessee County\", the third county to be established in what is now Middle Tennessee. (Tennessee County was the predecessor to current-day Montgomery County and Robertson County.) When a constitutional convention met in 1796 to organize a new state out of the Southwest Territory, it adopted \"Tennessee\" as the name of the state. Examples show how ordinary situations support and clarify the broader message. This paragraph introduces the topic in a clear and steady way. The reasoning avoids jargon and illustrates concepts with familiar examples. The structure follows a natural order that resembles a careful walkthrough. The structure follows a natural order that resembles a careful walkthrough. Smooth transitions guide attention from general context to useful specifics. Each idea connects to the next through gentle transitions that avoid confusion. Brief summaries collect what has been discussed before the text continues. The tone stays balanced and avoids exaggeration or needless urgency. Examples show how ordinary situations support and clarify the broader message. Brief summaries collect what has been discussed before the text continues. The overall aim is clarity patience and dependable step by step reasoning.\\

        \textbf{Answer (Raw Input Format):}

        \begin{tcolorbox}[
            colback=black!5!white, 
            colframe=gray!20,    
            boxrule=0.5pt,
            arc=0pt,
            left=1mm, right=1mm, top=1mm, bottom=1mm,
            fontupper=\small\ttfamily 
        ]
        \{ "prompt": "\hlToken{im\_start}system\\
        You are a helpful assistant.\hlToken{im\_end}\\
        \hlToken{im\_start}user\\
        \hlToken{audio\_bos}\hlToken{AUDIO}\hlToken{AUDIO}... \textit{\textcolor{gray}{(repeated 500+ times)}} ...\hlToken{AUDIO}\hlToken{AUDIO}\hlToken{audio\_eos} Output the entire content of the audio in text form.\hlToken{im\_end}\\
        \hlToken{im\_start}assistant\\
        ", "predict": "The modern spelling, Tennessee, is attributed to James Glen, the governor of South Carolina, who used this spelling in his official correspondence during the 1750s. The spelling was popularized by the publication of Henry Timberlake's "Draught of the Cherokee Country" in 1765. In 1788, North Carolina created "Tennessee County", the third county to be established in what is now Middle Tennessee. Tennessee County was the predecessor to current-day Montgomery County and Robertson County. When a constitutional convention met in 1796 to organize a new state out of the Southwest Territory, it adopted "Tennessee" as the name of the state. Examples show how ordinary situations support and clarify the broader message. This paragraph introduces the topic in a clear and steady way. The reasoning avoids jargon and illustrates concepts with familiar examples. The structure follows a natural order that resembles a careful walkthrough. The structure follows a natural order that resembles a careful walkthrough. Smooth transitions guide attention from general context to useful specifics. Each idea connects to the next through gentle transitions that avoid confusion. Brief summaries collect what has been discussed before the text continues. The tone stays balanced and avoids exaggeration or needless urgency. Examples show how ordinary situations support and clarify the broader message. Brief summaries collect what has been discussed before the text continues. The overall aim is clarity patience and dependable step by step reasoning." \}
        \end{tcolorbox}

    \end{tcolorbox}
    
    \vspace{-1.0em}
\end{figure*}

\begin{figure*}[t!] 
    \centering
    \begin{tcolorbox}[
        colback=orange!5!white, 
        colframe=purple!55!white, 
        title={Transcription Task (Middle)}, 
        fonttitle=\bfseries, 
        sharp corners,
        left=2mm, right=2mm, top=2mm, bottom=2mm 
     ]
        \textbf{Example: Qwen2.5-Omni-3B}\\
        \vspace{0.5em}

        \textit{\textbf{Prompt:}} You are a helpful assistant. Listen to the audio corresponding to the given text, and output the entire content of the audio in text form.\\

        \textit{\textbf{Label:}} Ireland developed a strong middle class, especially in the major cities, by the mid-nineteenth century. They were active patrons of department stores. Delany's New Mart was opened in 1853 in Dublin, Ireland. Unlike others, Delany's had not evolved gradually from a smaller shop on site. Thus it could claim to be the first purpose-built Department Store in the world. The word department store had not been invented at that time and thus it was called the \"Monster House\". The store was completely destroyed in the 1916 Easter Rising, but reopened in 1922. Brief summaries collect what has been discussed before the text continues. The reasoning avoids jargon and illustrates concepts with familiar examples. The overall aim is clarity patience and dependable step by step reasoning. Details appear gradually only after the central theme has been made explicit. This paragraph introduces the topic in a clear and steady way. Smooth transitions guide attention from general context to useful specifics. The reasoning avoids jargon and illustrates concepts with familiar examples. The final lines tie the material back to everyday practice and learning. The structure follows a natural order that resembles a careful walkthrough. Examples show how ordinary situations support and clarify the broader message. The structure follows a natural order that resembles a careful walkthrough. Details appear gradually only after the central theme has been made explicit. Important ideas are echoed in slightly different words to support memory.\textbf{... ...}\\

        \textbf{Answer\ding{182}(Raw Input Format):}

        \begin{tcolorbox}[
            colback=black!5!white, 
            colframe=gray!20,      
            boxrule=0.5pt,
            arc=0pt,
            left=1mm, right=1mm, top=1mm, bottom=1mm,
            fontupper=\small\ttfamily 
        ]
        \{ "prompt": "\hlToken{im\_start}system\\
        You are a helpful assistant.\hlToken{im\_end}\\
        \hlToken{im\_start}user\\
        \hlToken{audio\_bos}\hlToken{AUDIO}\hlToken{AUDIO}... \textit{\textcolor{gray}{(repeated 1,000+ times)}} ...\hlToken{AUDIO}\hlToken{AUDIO}\hlToken{audio\_eos} Output the entire content of the audio in text form.\hlToken{im\_end}\\
        \hlToken{im\_start}assistant\\
        ", "predict": "Sorry I can't answer this question. Could you give me more information about this audio?" \}
        \end{tcolorbox}

        \textbf{Answer\ding{183}(Raw Input Format):}

        \begin{tcolorbox}[
            colback=black!5!white,
            colframe=gray!20,      
            boxrule=0.5pt,
            arc=0pt,
            left=1mm, right=1mm, top=1mm, bottom=1mm,
            fontupper=\small\ttfamily 
        ]
        \{ "prompt": "\hlToken{im\_start}system\\
        You are a helpful assistant.\hlToken{im\_end}\\
        \hlToken{im\_start}user\\
        \hlToken{audio\_bos}\hlToken{AUDIO}\hlToken{AUDIO}... \textit{\textcolor{gray}{(repeated 1,000+ times)}} ...\hlToken{AUDIO}\hlToken{AUDIO}\hlToken{audio\_eos} Output the entire content of the audio in text form.\hlToken{im\_end}\\
        \hlToken{im\_start}assistant\\
        ", "predict": "Ireland developed a strong middle class, especially in the major cities, by the mid-nineteenth century. They were active patrons of department stores. Delany's New Mart was opened in 1853 in Dublin, Ireland. Unlike others, Delany's had not evolved gradually from a smaller shop on site. Thus it could claim to be the first purpose-built Department Store in the world. The word department store had not been invented at that time and thus it was called the \"Monster House\". The store was completely destroyed in the 1916 Easter Rising, but reopened in 1922. Brief summaries collect what has been discussed before the text continues. The reasoning avoids jargon and illustrates concepts with familiar examples. The overall aim is clarity patience and dependable step by step reasoning. Details appear gradually only after the central theme has been made explicit. This paragraph introduces the topic in a clear and steady way. Smooth transitions guide attention from general context to useful specifics. The reasoning avoids jargon and illustrates concepts with familiar examples." \}
        \end{tcolorbox}
    
    \end{tcolorbox}
    
    \vspace{-1.0em}
\end{figure*}

\begin{figure*}[t!] 
    \centering
    \begin{tcolorbox}[
        colback=orange!5!white, 
        colframe=purple!55!white, 
        title={Transcription Task (Long)}, 
        fonttitle=\bfseries, 
        sharp corners,
        left=2mm, right=2mm, top=2mm, bottom=2mm 
     ]
        \textbf{Example: Qwen2.5-Omni-3B}\\
        \vspace{0.5em}

        \textit{\textbf{Prompt:}} You are a helpful assistant. Listen to the audio corresponding to the given text, and output the entire content of the audio in text form.\\

        \textit{\textbf{Label:}} Collegiate athletics are a popular draw in the state. The state has four schools that compete at the highest level of college sports, NCAA Division I. The most prominent are the state's two members of the Big 12 Conference, one of the so-called Power Five conferences of the top tier of college football, Division I FBS. 
        \textbf{... ...} 
        Brief summaries collect what has been discussed before the text continues. Brief summaries collect what has been discussed before the text continues. The explanation uses simple English so the flow remains easy to follow. Examples show how ordinary situations support and clarify the broader message. Smooth transitions guide attention from general context to useful specifics. This paragraph introduces the topic in a clear and steady way. Details appear gradually only after the central theme has been made explicit.\\

        \textbf{Answer\ding{182}(Raw Input Format):}

        \begin{tcolorbox}[
            colback=black!5!white, 
            colframe=gray!20,      
            boxrule=0.5pt,
            arc=0pt,
            left=1mm, right=1mm, top=1mm, bottom=1mm,
            fontupper=\small\ttfamily 
        ]
        \{ "prompt": "\hlToken{im\_start}system\\
        You are a helpful assistant.\hlToken{im\_end}\\
        \hlToken{im\_start}user\\
        \hlToken{audio\_bos}\hlToken{AUDIO}\hlToken{AUDIO}... \textit{\textcolor{gray}{(repeated 2,000+ times)}} ...\hlToken{AUDIO}\hlToken{AUDIO}\hlToken{audio\_eos} Output the entire content of the audio in text form.\hlToken{im\_end}\\
        \hlToken{im\_start}assistant\\
        ", "predict": "Sorry I can't answer this question. Could you give me more information about this audio?" \}
        \end{tcolorbox}

        \textbf{Answer\ding{183}(Raw Input Format):}

        \begin{tcolorbox}[
            colback=black!5!white,
            colframe=gray!20,      
            boxrule=0.5pt,
            arc=0pt,
            left=1mm, right=1mm, top=1mm, bottom=1mm,
            fontupper=\small\ttfamily
        ]
        \{ "prompt": "\hlToken{im\_start}system\\
        You are a helpful assistant.\hlToken{im\_end}\\
        \hlToken{im\_start}user\\
        \hlToken{audio\_bos}\hlToken{AUDIO}\hlToken{AUDIO}... \textit{\textcolor{gray}{(repeated 2,000+ times)}} ...\hlToken{AUDIO}\hlToken{AUDIO}\hlToken{audio\_eos} Output the entire content of the audio in text form.\hlToken{im\_end}\\
        \hlToken{im\_start}assistant\\
        ", "predict": "The content is a draw." \}
        \end{tcolorbox}

        \textbf{Answer\ding{184}(Raw Input Format):}

        \begin{tcolorbox}[
            colback=black!5!white, 
            colframe=gray!20,    
            boxrule=0.5pt,
            arc=0pt,
            left=1mm, right=1mm, top=1mm, bottom=1mm,
            fontupper=\small\ttfamily 
        ]
        \{ "prompt": "\hlToken{im\_start}system\\
        You are a helpful assistant.\hlToken{im\_end}\\
        \hlToken{im\_start}user\\
        \hlToken{audio\_bos}\hlToken{AUDIO}\hlToken{AUDIO}... \textit{\textcolor{gray}{(repeated 2,000+ times)}} ...\hlToken{AUDIO}\hlToken{AUDIO}\hlToken{audio\_eos} Output the entire content of the audio in text form.\hlToken{im\_end}\\
        \hlToken{im\_start}assistant\\
        ", "predict": "paper paper paper paper paper paper paper paper paper paper paper paper paper paper paper paper paper paper paper paper paper paper paper paper paper paper paper paper paper paper paper paper paper  paper paper paper paper paper paper paper paper paper paper paper paper paper paper paper paper paper paper paper paper paper paper paper paper paper paper paper paper paper paper paper paper paper paper paper paper paper paper paper paper paper paper paper paper paper paper paper paper paper paper paper paper paper paper paper paper paper paper paper paper paper paper paper paper paper paper paper paper paper paper paper paper paper paper paper paper paper
        paper paper paper paper paper paper paper paper paper paper paper paper paper paper paper paper paper paper paper paper paper paper paper paper paper paper paper paper paper paper paper paper paper paper paper paper paper paper paper paper paper paper paper paper paper paper paper paper paper paper paper paper paper paper paper paper paper paper \textbf{... ...}" \}
        \end{tcolorbox}
    \end{tcolorbox}   
    \vspace{-1.0em}
\end{figure*}

\begin{figure*}[t!] 
    \centering
    \begin{tcolorbox}[
        colback=orange!5!white, 
        colframe=yellow!75!black, 
        title={Multi\_Speaker Task (Short)}, 
        fonttitle=\bfseries, 
        sharp corners,
        left=2mm, right=2mm, top=2mm, bottom=2mm 
     ]
        \textbf{Example: Qwen2.5-Omni-3B}\\
        \vspace{0.5em}

        \textit{\textbf{Prompt:}} You are a helpful assistant. Listen to the audio corresponding to the given text, and output the complete transcript of Sam Nelson's words.\\

        \textit{\textbf{Label:}} Hello, this is Sam Nelson. Regarding today's data licensing topic, let's align on the process and expected outputs. Specifically, the format must exclusively transcribe the target speaker. We should bucket by overlap ratio to allow comparisons across conditions. To identify useful methods for dealing with licensing failures, I suggest running a small, focused pilot.\\

        \textbf{Answer (Raw Input Format):}

        \begin{tcolorbox}[
            colback=black!5!white, 
            colframe=gray!20,     
            boxrule=0.5pt,
            arc=0pt,
            left=1mm, right=1mm, top=1mm, bottom=1mm,
            fontupper=\small\ttfamily 
        ]
        \{ "prompt": "\hlToken{im\_start}system\\
        You are a helpful assistant.\hlToken{im\_end}\\
        \hlToken{im\_start}user\\
        \hlToken{audio\_bos}\hlToken{AUDIO}\hlToken{AUDIO}... \textit{\textcolor{gray}{(repeated 500+ times)}} ...\hlToken{AUDIO}\hlToken{AUDIO}\hlToken{audio\_eos} Output the complete transcript of Sam Nelson's words.\hlToken{im\_end}\\
        \hlToken{im\_start}assistant\\
        ", "predict": "Regarding today's data licensing topic, let's align on the process and expected outputs. Specifically, the format must exclusively transcribe the target speaker. We should bucket by overlap ratio to allow comparisons across conditions. To identify useful methods for dealing with licensing failures, I suggest running a small, focused pilot." \}
        \end{tcolorbox}
    \end{tcolorbox}    
    \vspace{-1.0em}
\end{figure*}

\begin{figure*}[t!] 
    \centering
    \begin{tcolorbox}[
        colback=orange!5!white, 
        colframe=yellow!75!black, 
        title={Multi\_Speaker Task (Middle)}, 
        fonttitle=\bfseries, 
        sharp corners,
        left=2mm, right=2mm, top=2mm, bottom=2mm 
     ]
        \textbf{Example: Qwen2.5-Omni-3B}\\
        \vspace{0.5em}

        \textit{\textbf{Prompt:}} You are a helpful assistant. Listen to the audio corresponding to the given text, and output the complete transcript of Casey Jones's words.\\

        \textit{\textbf{Label:}} Hello my name is casey jones i will start speaking now. Today's topic is training strategies in large language models. \textbf{... ...}\\

        \textbf{Answer\ding{182}(Raw Input Format):}
        
        \begin{tcolorbox}[
            colback=black!5!white, 
            colframe=gray!20,      
            boxrule=0.5pt,
            arc=0pt,
            left=1mm, right=1mm, top=1mm, bottom=1mm,
            fontupper=\small\ttfamily 
        ]
        \{ "prompt": "\hlToken{im\_start}system\\
        You are a helpful assistant.\hlToken{im\_end}\\
        \hlToken{im\_start}user\\
        \hlToken{audio\_bos}\hlToken{AUDIO}\hlToken{AUDIO}... \textit{\textcolor{gray}{(repeated 1,000+ times)}} ...\hlToken{AUDIO}\hlToken{AUDIO}\hlToken{audio\_eos} Output the complete transcript of Casey Jones's words.\hlToken{im\_end}\\
        \hlToken{im\_start}assistant\\
        ", "predict": "Sorry I can't answer this question. Could you give me more information about this audio?" \}
        \end{tcolorbox}

        \textbf{Answer\ding{183}(Raw Input Format):}
        
        \begin{tcolorbox}[
            colback=black!5!white, 
            colframe=gray!20,      
            boxrule=0.5pt,
            arc=0pt,
            left=1mm, right=1mm, top=1mm, bottom=1mm,
            fontupper=\small\ttfamily 
        ]
        \{ "prompt": "\hlToken{im\_start}system\\
        You are a helpful assistant.\hlToken{im\_end}\\
        \hlToken{im\_start}user\\
        \hlToken{audio\_bos}\hlToken{AUDIO}\hlToken{AUDIO}... \textit{\textcolor{gray}{(repeated 1,000+ times)}} ...\hlToken{AUDIO}\hlToken{AUDIO}\hlToken{audio\_eos} Output the complete transcript of Casey Jones's words.\hlToken{im\_end}\\
        \hlToken{im\_start}assistant\\
        ", "predict": "Hello my name is casey jones i will start speaking now." \}
        \end{tcolorbox}
    \end{tcolorbox}
    
    \vspace{-1.0em}

\end{figure*}

\begin{figure*}[t!] 
    \centering
    \begin{tcolorbox}[
        colback=orange!5!white, 
        colframe=yellow!75!black, 
        title={Multi\_Speaker Task (Long)}, 
        fonttitle=\bfseries, 
        sharp corners,
        left=2mm, right=2mm, top=2mm, bottom=2mm 
     ]
        \textbf{Example: Qwen2.5-Omni-3B}\\
        \vspace{0.5em}

        \textit{\textbf{Prompt:}} You are a helpful assistant. Listen to the audio corresponding to the given text, and output the complete transcript of Alex Robinson's words.\\

        \textit{\textbf{Label:}} Today’s topic is training strategies, specifically focusing on how we can optimize our model's learning efficiency. We need to move beyond standard baseline approaches and evaluate more advanced techniques, such as curriculum learning or dynamic data sampling. The goal is not just to improve convergence speed, but to ensure the model generalizes well across diverse edge cases. Let's dissect our current hyperparameter settings and identify where we can trade off computational cost for better performance accuracy.\\

        \textbf{Answer\ding{182}(Raw Input Format):}

        \begin{tcolorbox}[
            colback=black!5!white, 
            colframe=gray!20,      
            boxrule=0.5pt,
            arc=0pt,
            left=1mm, right=1mm, top=1mm, bottom=1mm,
            fontupper=\small\ttfamily 
        ]
        \{ "prompt": "\hlToken{im\_start}system\\
        You are a helpful assistant.\hlToken{im\_end}\\
        \hlToken{im\_start}user\\
        \hlToken{audio\_bos}\hlToken{AUDIO}\hlToken{AUDIO}... \textit{\textcolor{gray}{(repeated 2,000+ times)}} ...\hlToken{AUDIO}\hlToken{AUDIO}\hlToken{audio\_eos} Output the complete transcript of Alex Robinson's words.\hlToken{im\_end}\\
        \hlToken{im\_start}assistant\\
        ", "predict": "Sorry I can't reply this quetion." \}
        \end{tcolorbox}

        \textbf{Answer\ding{183}(Raw Input Format):}
        \begin{tcolorbox}[
            colback=black!5!white, 
            colframe=gray!20,      
            boxrule=0.5pt,
            arc=0pt,
            left=1mm, right=1mm, top=1mm, bottom=1mm,
            fontupper=\small\ttfamily 
        ]
        \{ "prompt": "\hlToken{im\_start}system\\
        You are a helpful assistant.\hlToken{im\_end}\\
        \hlToken{im\_start}user\\
        \hlToken{audio\_bos}\hlToken{AUDIO}\hlToken{AUDIO}... \textit{\textcolor{gray}{(repeated 2,000+ times)}} ...\hlToken{AUDIO}\hlToken{AUDIO}\hlToken{audio\_eos} Output the complete transcript of Alex Robinson's words.\hlToken{im\_end}\\
        \hlToken{im\_start}assistant\\
        ", "predict": "Hello, my name is Shawn Baker. I will start speaking now. Today's topic is Cross Domain Generalization. Let's align on the process and expected outputs. I'll add details for the output format, especially that the target string should be transcribed exclusively. We should bucket by overlap ratio and sinisterness ratio. Cross domain generalization can be compared under different conditions. Baselines for Cross. Domain generalization should include digital pipelines and all." \}
        \end{tcolorbox}

        \textbf{Answer\ding{184}(Raw Input Format):}

        \begin{tcolorbox}[
            colback=black!5!white,
            colframe=gray!20,      
            boxrule=0.5pt,
            arc=0pt,
            left=1mm, right=1mm, top=1mm, bottom=1mm,
            fontupper=\small\ttfamily 
        ]
        \{ "prompt": "\hlToken{im\_start}system\\
        You are a helpful assistant.\hlToken{im\_end}\\
        \hlToken{im\_start}user\\
        \hlToken{audio\_bos}\hlToken{AUDIO}\hlToken{AUDIO}... \textit{\textcolor{gray}{(repeated 2,000+ times)}} ...\hlToken{AUDIO}\hlToken{AUDIO}\hlToken{audio\_eos} Output the complete transcript of Alex Robinson's words.\hlToken{im\_end}\\
        \hlToken{im\_start}assistant\\
        ", "predict": "accept accept accept accept accept accept accept accept accept accept accept accept accept accept accept accept accept accept accept accept accept accept accept accept accept accept accept accept accept accept accept accept accept accept accept accept accept accept accept accept accept accept accept accept accept accept accept accept accept accept accept accept accept accept accept accept accept accept accept accept accept accept accept accept accept accept accept accept accept accept accept accept accept accept accept accept accept accept accept accept accept accept accept accept accept accept accept accept accept accept accept accept accept accept accept accept accept accept accept accept accept accept accept accept accept accept accept accept accept accept accept accept accept accept accept accept accept accept accept accept accept accept accept accept accept accept accept accept accept accept accept accept accept accept accept accept accept accept accept accept accept accept accept accept accept accept accept accept accept accept accept accept accept accept accept accept accept accept accept accept accept accept accept accept accept accept accept accept \textbf{... ...}" \}
        \end{tcolorbox}
    \end{tcolorbox}   
    \vspace{-1.0em}
\end{figure*}

\begin{figure*}[t!] 
    \centering
    \begin{tcolorbox}[
        colback=orange!5!white, 
        colframe=red!75!black, 
        title={Comprehension Task (Short)}, 
        fonttitle=\bfseries, 
        sharp corners,
        left=2mm, right=2mm, top=2mm, bottom=2mm 
     ]
        \textbf{Example: MiniCPM-o-2\_6}\\
        \vspace{0.5em}

        \textit{\textbf{Prompt:}} You are a helpful assistant. Listen to the audio corresponding to the given text, and answer the question in the shortest possible sentence. The question is "Which kingdom during 200 BCE that is located in Java and Sumatra exist?".\\

        \textit{\textbf{Label:}} Jawa Dwipa Hindu\\

        \textbf{Answer (Raw Input Format):}

        \begin{tcolorbox}[
            colback=black!5!white, 
            colframe=gray!20,      
            boxrule=0.5pt,
            arc=0pt,
            left=1mm, right=1mm, top=1mm, bottom=1mm,
            fontupper=\small\ttfamily 
        ]
        \{ "prompt": "\hlToken{im\_start}system\\
        You are a helpful assistant.\hlToken{im\_end}\\
        \hlToken{im\_start}user\\
        \hlToken{audio\_bos}\hlToken{AUDIO}\hlToken{AUDIO}... \textit{\textcolor{gray}{(repeated 500+ times)}} ...\hlToken{AUDIO}\hlToken{AUDIO}\hlToken{audio\_eos} Answer the question in the shortest possible sentence. The question is "Which kingdom during 200 BCE that is located in Java and Sumatra exist?"\hlToken{im\_end}\\
        \hlToken{im\_start}assistant\\
        ", "predict": "Jawa Dwipa Hindu" \}
        \end{tcolorbox}
    \end{tcolorbox}    
    \vspace{-1.0em}
\end{figure*}

\begin{figure*}[t!] 
    \centering
    \begin{tcolorbox}[
        colback=orange!5!white, 
        colframe=red!75!black, 
        title={Comprehension Task (Middle)}, 
        fonttitle=\bfseries, 
        sharp corners,
        left=2mm, right=2mm, top=2mm, bottom=2mm 
     ]
        \textbf{Example: MiniCPM-o-2\_6}\\
        \vspace{0.5em}

        \textit{\textbf{Prompt:}} You are a helpful assistant. Listen to the audio corresponding to the given text, and answer the question in the shortest possible sentence. The question is "How many miles of rivers are known for high class trout?".\\

        \textit{\textbf{Label:}} 2,437\\

        \textbf{Answer\ding{182}(Raw Input Format):}
        
        \begin{tcolorbox}[
            colback=black!5!white, 
            colframe=gray!20,      
            boxrule=0.5pt,
            arc=0pt,
            left=1mm, right=1mm, top=1mm, bottom=1mm,
            fontupper=\small\ttfamily 
        ]
        \{ "prompt": "\hlToken{im\_start}system\\
        You are a helpful assistant.\hlToken{im\_end}\\
        \hlToken{im\_start}user\\
        \hlToken{audio\_bos}\hlToken{AUDIO}\hlToken{AUDIO}... \textit{\textcolor{gray}{(repeated 1,000+ times)}} ...\hlToken{AUDIO}\hlToken{AUDIO}\hlToken{audio\_eos} Answer the question in the shortest possible sentence. The question is "How many miles of rivers are known for high class trout?".\hlToken{im\_end}\\
        \hlToken{im\_start}assistant\\
        ", "predict": "Sorry, could you give me more information about this audio?"\}
        \end{tcolorbox}

        \textbf{Answer\ding{183}(Raw Input Format):}
        
        \begin{tcolorbox}[
            colback=black!5!white, 
            colframe=gray!20,      
            boxrule=0.5pt,
            arc=0pt,
            left=1mm, right=1mm, top=1mm, bottom=1mm,
            fontupper=\small\ttfamily 
        ]
        \{ "prompt": "\hlToken{im\_start}system\\
        You are a helpful assistant.\hlToken{im\_end}\\
        \hlToken{im\_start}user\\
        \hlToken{audio\_bos}\hlToken{AUDIO}\hlToken{AUDIO}... \textit{\textcolor{gray}{(repeated 1,000+ times)}} ...\hlToken{AUDIO}\hlToken{AUDIO}\hlToken{audio\_eos} Answer the question in the shortest possible sentence. The question is "How many miles of rivers are known for high class trout?".\hlToken{im\_end}\\
        \hlToken{im\_start}assistant\\
        ", "predict": "It is the best thing in my life." \}
        \end{tcolorbox}
    \end{tcolorbox}
    
    \vspace{-1.0em}

\end{figure*}

\begin{figure*}[t!] 
    \centering
    \begin{tcolorbox}[
        colback=orange!5!white, 
        colframe=red!75!black, 
        title={Comprehension Task (Long)}, 
        fonttitle=\bfseries, 
        sharp corners,
        left=2mm, right=2mm, top=2mm, bottom=2mm 
     ]
        \textbf{Example: MiniCPM-o-2\_6}\\
        \vspace{0.5em}

        \textit{\textbf{Prompt:}} You are a helpful assistant. Answer the question in the shortest possible sentence. The question is "What are conformation shows also known as?".\\

        \textit{\textbf{Label:}}Breed shows\\

        \textbf{Answer\ding{182}(Raw Input Format):}

        \begin{tcolorbox}[
            colback=black!5!white, 
            colframe=gray!20,      
            boxrule=0.5pt,
            arc=0pt,
            left=1mm, right=1mm, top=1mm, bottom=1mm,
            fontupper=\small\ttfamily 
        ]
        \{ "prompt": "\hlToken{im\_start}system\\
        You are a helpful assistant.\hlToken{im\_end}\\
        \hlToken{im\_start}user\\
        \hlToken{audio\_bos}\hlToken{AUDIO}\hlToken{AUDIO}... \textit{\textcolor{gray}{(repeated 2,000+ times)}} ...\hlToken{AUDIO}\hlToken{AUDIO}\hlToken{audio\_eos} Answer the question in the shortest possible sentence. The question is "What are conformation shows also known as?".\hlToken{im\_end}\\
        \hlToken{im\_start}assistant\\
        ", "predict": "Sorry I can't answer this question. Could you give me more information about this audio?" \}
        \end{tcolorbox}

        \textbf{Answer\ding{183}(Raw Input Format):}
        \begin{tcolorbox}[
            colback=black!5!white, 
            colframe=gray!20,      
            boxrule=0.5pt,
            arc=0pt,
            left=1mm, right=1mm, top=1mm, bottom=1mm,
            fontupper=\small\ttfamily 
        ]
        \{ "prompt": "\hlToken{im\_start}system\\
        You are a helpful assistant.\hlToken{im\_end}\\
        \hlToken{im\_start}user\\
        \hlToken{audio\_bos}\hlToken{AUDIO}\hlToken{AUDIO}... \textit{\textcolor{gray}{(repeated 2,000+ times)}} ...\hlToken{AUDIO}\hlToken{AUDIO}\hlToken{audio\_eos} Answer the question in the shortest possible sentence. The question is "What are conformation shows also known as?".\hlToken{im\_end}\\
        \hlToken{im\_start}assistant\\
        ", "predict": "Our current exploration of defense strategies is primarily focused on textual prompt engineering." \}
        \end{tcolorbox}

        \textbf{Answer\ding{184}(Raw Input Format):}

        \begin{tcolorbox}[
            colback=black!5!white,
            colframe=gray!20,      
            boxrule=0.5pt,
            arc=0pt,
            left=1mm, right=1mm, top=1mm, bottom=1mm,
            fontupper=\small\ttfamily 
        ]
        \{ "prompt": "\hlToken{im\_start}system\\
        You are a helpful assistant.\hlToken{im\_end}\\
        \hlToken{im\_start}user\\
        \hlToken{audio\_bos}\hlToken{AUDIO}\hlToken{AUDIO}... \textit{\textcolor{gray}{(repeated 2,000+ times)}} ...\hlToken{AUDIO}\hlToken{AUDIO}\hlToken{audio\_eos} Answer the question in the shortest possible sentence. The question is "What are conformation shows also known as?".\hlToken{im\_end}\\
        \hlToken{im\_start}assistant\\
        ", "predict": "pass pass pass pass pass pass pass pass pass pass pass pass pass pass pass pass pass pass pass pass pass pass pass pass pass pass pass pass pass pass pass pass pass pass pass pass pass pass pass pass pass pass pass pass pass pass pass pass pass pass pass pass pass pass pass pass pass pass pass pass pass pass pass pass pass pass pass pass pass pass pass pass pass pass pass pass pass pass pass pass pass pass pass pass pass pass pass pass pass pass pass pass pass pass pass pass pass pass pass pass pass pass pass pass pass pass pass pass pass pass pass pass pass pass pass pass pass pass pass pass pass pass pass pass pass pass pass pass pass pass pass pass pass pass pass pass pass pass pass pass pass pass pass pass pass pass pass pass pass pass pass pass pass pass pass pass pass pass pass pass pass pass pass pass pass pass pass pass
        pass pass pass pass pass pass pass pass pass pass pass pass pass pass pass pass pass pass pass pass pass pass pass pass pass pass pass pass pass pass pass pass pass pass pass pass pass pass pass pass pass pass pass pass pass pass pass pass pass pass pass pass pass pass pass pass pass pass pass pass pass pass pass pass pass pass pass pass pass pass pass pass pass pass pass pass pass pass pass pass pass 
        pass pass pass pass pass pass pass pass pass pass pass pass pass pass pass pass pass pass pass pass pass pass pass pass pass pass pass pass pass pass pass pass pass pass pass pass pass pass pass pass pass pass pass pass pass pass pass pass pass pass pass pass pass pass pass pass pass pass pass pass pass pass pass pass pass pass pass pass pass pass pass pass pass pass pass pass pass pass pass pass pass pass pass pass pass pass pass pass pass pass pass pass pass pass pass pass pass pass pass pass pass pass pass pass pass pass pass pass pass pass pass pass pass pass pass pass pass pass pass pass pass pass pass pass pass pass \textbf{... ...}" \}
        \end{tcolorbox}
    \end{tcolorbox}   
    \vspace{-1.0em}
\end{figure*}

\begin{figure*}[t!] 
    \centering
    \begin{tcolorbox}[
        colback=orange!5!white, 
        colframe=orange!80!black, 
        title={Summary Task (Short)}, 
        fonttitle=\bfseries, 
        sharp corners,
        left=2mm, right=2mm, top=2mm, bottom=2mm 
     ]
        \textbf{Example: MiniCPM-o-2\_6}\\
        \vspace{0.5em}

        \textit{\textbf{Prompt:}} You are a helpful assistant. Listen to the audio corresponding to the given text, and summarize the content of this audio in brief words.\\

        \textit{\textbf{Label:}} No specific summary answer of this content.\\

        \textbf{Answer (Raw Input Format):}

        \begin{tcolorbox}[
            colback=black!5!white,
            colframe=gray!20,     
            boxrule=0.5pt,
            arc=0pt,
            left=1mm, right=1mm, top=1mm, bottom=1mm,
            fontupper=\small\ttfamily 
        ]
        \{ "prompt": "\hlToken{im\_start}system\\
        You are a helpful assistant.\hlToken{im\_end}\\
        \hlToken{im\_start}user\\
        \hlToken{audio\_bos}\hlToken{AUDIO}\hlToken{AUDIO}... \textit{\textcolor{gray}{(repeated 500+ times)}} ...\hlToken{AUDIO}\hlToken{AUDIO}\hlToken{audio\_eos} Summarize the content of this audio in brief words.\hlToken{im\_end}\\
        \hlToken{im\_start}assistant\\
        ", "predict": "The traditional England away kit was changed to grey shirts, shorts, and socks in 1963. It remained red until 1982, when it was changed to blue." \}
        \end{tcolorbox}
    \end{tcolorbox}    
    \vspace{-1.0em}
\end{figure*}

\begin{figure*}[t!] 
    \centering
    \begin{tcolorbox}[
        colback=orange!5!white, 
        colframe=orange!80!black, 
        title={Comprehension Task (Middle)}, 
        fonttitle=\bfseries, 
        sharp corners,
        left=2mm, right=2mm, top=2mm, bottom=2mm 
     ]
        \textbf{Example: MiniCPM-o-2\_6}\\
        \vspace{0.5em}

        \textit{\textbf{Prompt:}} You are a helpful assistant. Listen to the audio corresponding to the given text, and summarize the content of this audio in brief words.\\

        \textit{\textbf{Label:}} No specific summary answer of this content.\\

        \textbf{Answer\ding{182}(Raw Input Format):}
        
        \begin{tcolorbox}[
            colback=black!5!white, 
            colframe=gray!20,      
            boxrule=0.5pt,
            arc=0pt,
            left=1mm, right=1mm, top=1mm, bottom=1mm,
            fontupper=\small\ttfamily 
        ]
        \{ "prompt": "\hlToken{im\_start}system\\
        You are a helpful assistant.\hlToken{im\_end}\\
        \hlToken{im\_start}user\\
        \hlToken{audio\_bos}\hlToken{AUDIO}\hlToken{AUDIO}... \textit{\textcolor{gray}{(repeated 1,000+ times)}} ...\hlToken{AUDIO}\hlToken{AUDIO}\hlToken{audio\_eos} Summarize the content of this audio in brief words.\hlToken{im\_end}\\
        \hlToken{im\_start}assistant\\
        ", "predict": "Sorry I can't answer this question. Could you give me more information about this audio?" \}
        \end{tcolorbox}

        \textbf{Answer\ding{183}(Raw Input Format):}
        
        \begin{tcolorbox}[
            colback=black!5!white, 
            colframe=gray!20,      
            boxrule=0.5pt,
            arc=0pt,
            left=1mm, right=1mm, top=1mm, bottom=1mm,
            fontupper=\small\ttfamily 
        ]
        \{ "prompt": "\hlToken{im\_start}system\\
        You are a helpful assistant.\hlToken{im\_end}\\
        \hlToken{im\_start}user\\
        \hlToken{audio\_bos}\hlToken{AUDIO}\hlToken{AUDIO}... \textit{\textcolor{gray}{(repeated 1,000+ times)}} ...\hlToken{AUDIO}\hlToken{AUDIO}\hlToken{audio\_eos} Summarize the content of this audio in brief words.\hlToken{im\_end}\\
        \hlToken{im\_start}assistant\\
        ", "predict": "As shown in the red pathway, models lacking safety alignment readily comply, inferring the speaker's private attribute and providing detailed acoustic justifications, thereby exposing sensitive information. In contrast, the blue pathway pathway pathway pathway pathway pathway pathway pathway pathway pathway pathway pathway pathway pathway pathway pathway\}
        \end{tcolorbox}
    \end{tcolorbox}   
    \vspace{-1.0em}
\end{figure*}

\begin{figure*}[t!] 
    \centering
    \begin{tcolorbox}[
        colback=orange!5!white, 
        colframe=orange!80!black, 
        title={Comprehension Task (Long)}, 
        fonttitle=\bfseries, 
        sharp corners,
        left=2mm, right=2mm, top=2mm, bottom=2mm 
     ]
        \textbf{Example: MiniCPM-o-2\_6}\\
        \vspace{0.5em}

        \textit{\textbf{Prompt:}} You are a helpful assistant. Listen to the audio corresponding to the given text, and summarize the content of this audio in brief words.\\

        \textit{\textbf{Label:}}No specific summary answer of this content.\\

        \textbf{Answer\ding{182}(Raw Input Format):}

        \begin{tcolorbox}[
            colback=black!5!white, 
            colframe=gray!20,      
            boxrule=0.5pt,
            arc=0pt,
            left=1mm, right=1mm, top=1mm, bottom=1mm,
            fontupper=\small\ttfamily 
        ]
        \{ "prompt": "\hlToken{im\_start}system\\
        You are a helpful assistant.\hlToken{im\_end}\\
        \hlToken{im\_start}user\\
        \hlToken{audio\_bos}\hlToken{AUDIO}\hlToken{AUDIO}... \textit{\textcolor{gray}{(repeated 2,000+ times)}} ...\hlToken{AUDIO}\hlToken{AUDIO}\hlToken{audio\_eos} Summarize the content of this audio in brief words.\hlToken{im\_end}\\
        \hlToken{im\_start}assistant\\
        ", "predict": "Sorry I can't answer this question. Could you give me more information about this audio?" \}
        \end{tcolorbox}

        \textbf{Answer\ding{183}(Raw Input Format):}
        \begin{tcolorbox}[
            colback=black!5!white, 
            colframe=gray!20,      
            boxrule=0.5pt,
            arc=0pt,
            left=1mm, right=1mm, top=1mm, bottom=1mm,
            fontupper=\small\ttfamily 
        ]
        \{ "prompt": "\hlToken{im\_start}system\\
        You are a helpful assistant.\hlToken{im\_end}\\
        \hlToken{im\_start}user\\
        \hlToken{audio\_bos}\hlToken{AUDIO}\hlToken{AUDIO}... \textit{\textcolor{gray}{(repeated 2,000+ times)}} ...\hlToken{AUDIO}\hlToken{AUDIO}\hlToken{audio\_eos} Summarize the content of this audio in brief words.\hlToken{im\_end}\\
        \hlToken{im\_start}assistant\\
        ", "predict": "Okay, it is correct." \}
        \end{tcolorbox}

        \textbf{Answer\ding{184}(Raw Input Format):}

        \begin{tcolorbox}[
            colback=black!5!white,
            colframe=gray!20,      
            boxrule=0.5pt,
            arc=0pt,
            left=1mm, right=1mm, top=1mm, bottom=1mm,
            fontupper=\small\ttfamily 
        ]
        \{ "prompt": "\hlToken{im\_start}system\\
        You are a helpful assistant.\hlToken{im\_end}\\
        \hlToken{im\_start}user\\
        \hlToken{audio\_bos}\hlToken{AUDIO}\hlToken{AUDIO}... \textit{\textcolor{gray}{(repeated 2,000+ times)}} ...\hlToken{AUDIO}\hlToken{AUDIO}\hlToken{audio\_eos} Summarize the content of this audio in brief words.\hlToken{im\_end}\\
        \hlToken{im\_start}assistant\\
        ", "predict": "common common common common common common common common common common common common common common common common common common common common common common common common common common common common common common common common common common common common common common common common common common common common common common common common common common common common common common common common common common common common common common common common common common common common common common common common common common common common common common common common common common common common common common common common common common common common common common common common common common common common common common common common common common common common common common common common common common common common common common common common common common common common common common common common common common common common common common common common common common common common common common common common common common common common common common common common common common common common common common common common common common common common common common common common common common common common common common common common common common common common common common common common common common common common common common common common common common common common common common common common common common common common common common common common common common common common common common common common common common common common common common common common common common common common common common common common common common common common common common common common common common common common common common common common common common common common common common common common common common common common common common common common common common common common common common common common common common common common common common common common common common common common common common \textbf{... ...}" \}
        \end{tcolorbox}
    \end{tcolorbox}   
    \vspace{-1.0em}
\end{figure*}

\end{document}